\documentclass[pra, twocolumn, floatfix]{revtex4-2}
\usepackage{tikz}
\usetikzlibrary{arrows, shapes, decorations.markings, calc, intersections, backgrounds, shapes.misc, external, positioning, fit}
\usepackage{pgfplots}
\pgfplotsset{compat=newest}
\usepgfplotslibrary{fillbetween}

\usepackage{slashed}
\usepackage{graphicx}
\usepackage{subfigure}
\usepackage[colorlinks=true,linkcolor=blue,citecolor=blue,urlcolor=blue]{hyperref}
\usepackage{amsmath}
\usepackage{amsfonts}
\usepackage[none]{hyphenat}

\usepackage{txfonts}
\usepackage{times} 

\usepackage[british]{babel}

\definecolor{Cedar}{RGB}{182, 170, 167}
\definecolor{Concrete}{RGB}{179,189,177}
\definecolor{Sky}{RGB}{165, 200, 208}
\definecolor{Heather}{RGB}{203 168 177}
\definecolor{Blue}{RGB}{0, 174, 239}
\definecolor{Purple}{RGB}{104, 36, 109}
\definecolor{Red}{RGB}{190, 30, 45}
\definecolor{Gold}{RGB}{175, 169, 97}
\definecolor{Yellow}{RGB}{255 213 58}
\definecolor{Stone}{RGB}{218 205 162}
\definecolor{Green}{RGB}{0, 130, 0}

\newcommand{\pr}[1]{\ensuremath{\left[#1\right]}} 
\newcommand{\pc}[1]{\ensuremath{\left(#1\right)}} 
\newcommand{\px}[1]{\ensuremath{\left\lbrace#1\right\rbrace}} 
\newcommand{\bra}[1]{\ensuremath{\left\langle#1\right\vert}} 
\newcommand{\ket}[1]{\ensuremath{\left\vert#1\right\rangle}} 
\newcommand{\md}[1]{\ensuremath{\left\vert#1\right\vert}} 
\newcommand{\av}[1]{\ensuremath{\left\langle#1\right\rangle}}

\DeclareMathOperator{\hH}{\hat{H}}
\DeclareMathOperator{\ha}{\hat{a}}

\DeclareMathOperator{\hO}{\hat{O}}

\begin{document}
\title{Collective effects in the photon statistics of thermal atomic ensembles}

\author{Sofia Ribeiro}
\author{Thomas F. Cutler}
\author{Charles S. Adams}
\author{Simon A. Gardiner}
\affiliation{Joint Quantum Center (JQC) Durham--Newcastle, Department of Physics,  Durham University, South Road, Durham DH1 3LE, United Kingdom}

\begin{abstract}
We investigate the collective scattering of coherent light from a thermal alkali-metal vapor with temperatures ranging from 350 to 450~K, corresponding to average atomic spacings between $0.7 \lambda$ and $0.1 \lambda$. We develop a theoretical model treating the atomic ensemble as coherent, interacting, radiating dipoles. We show that the two-time second-order correlation function of a thermal ensemble can be described by an average of randomly positioned atomic pairs. Our model illustrates good qualitative agreement with the experimental results. Furthermore, we show how fine-tuning of the experimental parameters may make it possible to explore several photon statistics regimes. 
\end{abstract}

\maketitle
\section{Introduction}
Since the pioneering experiment carried out by Hanbury Brown and Twiss in 1956 \cite{Nature177.27(1956), Nature178.1046(1956)}, the study of photon statistics has been of great importance in understanding various phenomena in quantum optics. More recently, control of light at few-photon levels has become critical to quantum technologies, where different applications require different regimes. One such regime is characterized by strong anti-bunching, working towards the development of single-photon sources. Such sources have many applications in quantum computation, simulation and sensing \cite{NatPhot3.696(2009), RevScienInst82.071101(2011), OptPhotNews30.32(2019)}. Another regime is characterized by strong photon bunching, and sources displaying this characteristic have technological applications in imaging and interference experiments \cite{PRL74.3600(1995), PRA70.051802(R)(2004), PRA73.053802(2006), PRL119.263603(2017)}. 
A common way to characterize and classify a light field is by measuring its second-order correlation function, $g^{(2)} (\tau)$. Such studies have been made for atomic beams \cite{PRL16.1012(1966)}, cold atoms \cite{PRA53.3469(1996), OptLett29.2713(2004),OptExp18.6604(2010)}, thermal atomic vapor cells \cite{PRA93.043826(2016)}, and solid-state systems \cite{PRB78.153309(2008)}. 
Anti-bunched sources have sub-Poissonian statistics with $g^{(2)} (0) \ll 1$, while bunched sources have super-Poissonian statistics with $g^{(2)} (0) \gg 1$. An ideal thermal source has $g^{(2)} (0) = 2$, and such ideal or pseudo-thermal statistics have been demonstrated in spinning glass discs \cite{JOSA61.1307(1971), PRL116.050401(2016)}, cold atoms \cite{OptExp18.6604(2010), JPhysB49.025301(2016)}, and hot atomic vapors \cite{PRA93.043826(2016), NJP20.093002(2018), ScienRep8.10981(2018)}.
Conversely, recent efforts in search of strongly anti-bunched sources have culminated in such sub-Poissonian statistics being measured for a wide variety of sources, including cold atoms \cite{PRL92.213601(2004)}, Rydberg atoms \cite{Science336.887(2012)}, single ions \cite{NJP11.103004(2009)} and molecules \cite{PRL83.2722(1999), Science298.385(2002)}, quantum dots \cite{npjQuantInf(2018)}, solid-state sources such as nitrogen-vacancy centers (NVCs) \cite{PRL85.290(2000)},  and  thermal vapors~\cite{PRA87.053412(2013), PRL118.253602(2017), Science362.446(2018)}. 
However, there are very few demonstrations of both bunched and anti-bunched regimes in a precise and controlled manner \cite{Science298.385(2002)}. Ideally, a single system would be used to cross over between regimes through tuning experimental parameters. Many experimental teams have focused on trapped atoms to achieve such cross-over \cite{PRA101.023828(2020), PRL124.063603(2020)}. 
However, room temperature atomic vapor experiments have the advantage of being simpler, less expensive, and, most importantly, more compact and scalable. Producing a full many-body description of these systems, especially in the case of a dense atomic medium, nevertheless remains a challenge \cite{PRA93.043826(2016), PRA100.033833(2019), PRL122.183203(2019)}

Light propagation and light-matter interaction in dense atomic samples are rich areas of study. The interplay between matter and light gives rise to novel concepts, as introduced in Dicke's seminal work \cite{PhysRev93.99(1954)}. Important effects arise from the dipole--dipole interactions, such as collective level shifts and line broadening. As a consequence of the strong dipole--dipole interactions, the behavior of an ensemble of $N$ atoms cannot be described by summing the response of a single atom $N$ times.
Previous work has treated the subject of photon statistics for ensembles of two or more two-level atoms. In Ref.~\cite{PRA15.1613(1977)}, for example, the authors treated each atom as interacting with the source field independently of the other surrounding atoms and found substantial differences from the single atom treatment. The appearance of non-classical correlations in the radiation of two atoms that are coherently driven by a continuous laser source occurs even without any inter-atomic interaction \cite{PRA64.063801(2001)}.
Several theoretical approaches can be used to treat the system, including dipole--dipole interactions. In Ref.~\cite{PRA19.1132(1979)}, a dressed-atom approach to resonance fluorescence in intense laser fields is presented. It is found that the inclusion of dipole--dipole interactions changes the spectrum of the two-atom system considerably from that of a single atom. For a two-atom system, both photon bunching and anti-bunching can occur in the scattered radiation \cite{OptActa29.265(1982)}. The cooperative behavior and the dipole--dipole interaction are shown to act to diminish the photon anti-bunching \cite{PRA25.1528(1982)}, while squeezing can also occur in two-atom resonance fluorescence \cite{PRA29.2004(1984)}. 
In these previous works, the treatment of two-atom systems assumes that both atoms experience the same electromagnetic field, and their relative distance remains constant over the radiation process. Although these assumptions simplify both analytical and numerical treatments, they are not fulfilled in realistic experiments.
Even though numerous theoretical efforts have been made over the years, experimental demonstrations that explore the temporal photon correlations from a thermal vapor have been elusive. One study \cite{PRA93.043826(2016)} has investigated the photon statistics of a thermal vapor at different temperatures. However, the results presented are strongly influenced and complicated by re-scattering due to the high optical depth attained at higher temperatures in a conventional millimeter-scale vapor cell. The confinement regimes possible within nano-cells substantially eliminate the issue of re-scattering, and thus, in this paper, it is possible to experimentally access previously unstudied high-density regimes in the photon statistics of thermal vapors.

In this article, we study theoretically a dense thermal cloud of rubidium atoms confined in a vapor cell. We closely follow a methodology that includes dipole--dipole interactions and considers that the atoms experience different intensities and phases of the driving field, as described in Ref.~\cite{PRA52.636(1995)}.
Furthermore, in our study, we average the behavior of many possible inter-atomic separations between the atomic pairs. This is equivalent to studying the collective effects for an idealized thermal vapor consisting of a random distribution of atoms moving with different velocities. 
We discuss and illustrate how these theoretical predictions relate to experimental observations of a nanoscale thermal vapor's photon statistics with varying density, as reported in Ref.~\cite{WillThesis}.
The second-order correlation function gives insight into the quantum nature of the source, as has been probed experimentally. We show that by controlling the atomic density, driving, and detuning, it is possible to explore several regimes in the thermal vapor photon statistics. 

\begin{figure}
\centering
\includegraphics[width=\linewidth, keepaspectratio]{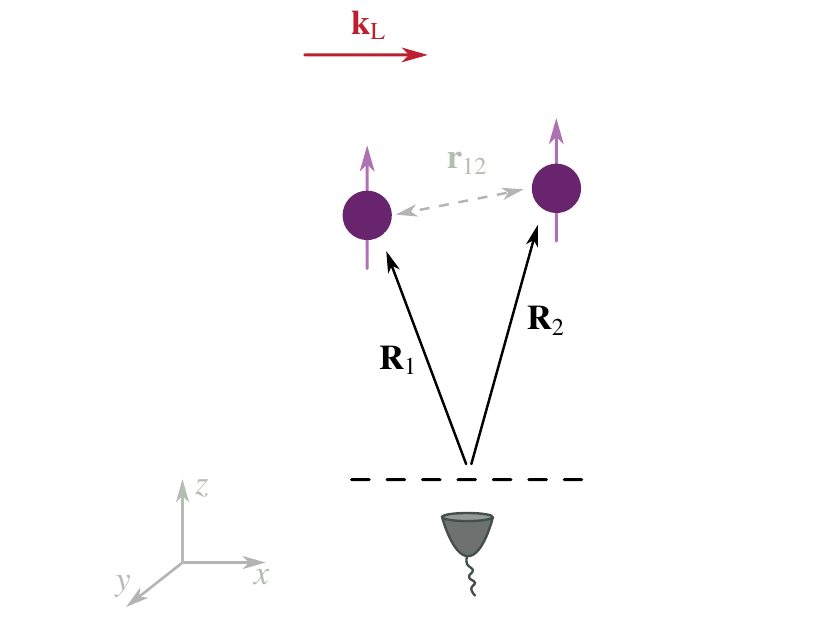}
\caption{(Color online) Scheme for the geometry for the observation of intensity correlations. The photon statistics are considered to be measured at a point in the far-field zone of the radiation emitted by the atomic system $\pc{R_{1,2} \gg r_{12}, \lambda,\; \hat{\mathbf{R}} =\hat{\mathbf{R}}_1 =\hat{\mathbf{R}}_2 }$. For simplicity, we consider the direction of observation along the $z$-axis, and the driving field $\mathbf{k}_\text{L} = \pc{k_\text{L},\, 0,\, 0}$.
\label{fig:ExpSetup}}
\end{figure}

\section{Theoretical model \label{sec:Theory}} 
%
\subsection{Two-atom master equation}
We wish to analyze the dynamics of a dense thermal atomic vapor. To investigate its photon statistics, we consider two two-level atoms, at fixed positions $\mathbf{r}_1$ and $\mathbf{r}_2$ (where we also define $\mathbf{r}_{12} = \mathbf{r}_1 - \mathbf{r}_2$), with dipole moment $\mathbf{d}_{eg}$ for ground state $\ket{g}$ and excited state $\ket{e}$, and transition frequency $\omega_0$. Note this is a simplification of real alkali-metal atoms which are multi-level, however it is possible to isolate a particular level using large magnetic fields \cite{OptLet40.4289(2015)}.  

The atoms are driven by an external laser field with wavevector $\mathbf{k}_\text{L}$. In a running-wave laser field, the coupling strength experienced by the $i{\text{th}}$ particle is described by $\Omega _i=\Omega_\text{R}  \exp \left(-\text{i} \mathbf{k}_\text{L} \cdot \mathbf{r}_i \right)$ where $\Omega_\text{R}$ is the maximum Rabi frequency and $\mathbf{r}_i$ is the atomic position vector. 
For an atomic vapor, loss of energy from an excited atom can occur via spontaneous emission or inelastic collisions. Elastic collisions may also occur, inducing only a change of phase of induced electronic oscillations. For the sake of simplicity, we assume that collisions between atoms are not important and the only dissipative terms are due to the spontaneous decays of the levels $\ket{e}_{1,2}$ at a rate $\Gamma$. 
The Hamiltonian $\hH$ that describes the system is composed of three terms: the unperturbed Hamiltonian of the atoms, the dipole--dipole interaction between the atoms, and the coupling between the driving field and the atoms. The dynamics of the system will be described by the reduced atomic density operator $\rho$. Following closely Refs.~\cite{PhysRep372.369(2002), PRA52.636(1995)}, the master equation can be written as 
\begin{align}
\frac{\partial \rho }{\partial t} = & -\text{i} \omega_0 \sum_{i=1}^2 \pr{ \sigma_i^z , \, \rho} 
-\frac{\text{i}}{2}  \sum_{i \neq j}  g_{ij} \pr{\sigma_i^+ \sigma_j^- + \text{H.c.} , \, \rho} \nonumber \\
&
+\frac{\text{i}}{2 } \sum_{i=1}^2 \pr{\Omega_i \, \sigma_i^+ \exp \pc{\text{i} \omega_\text{L} t} + \text{H.c.}, \, \rho} 
\nonumber \\
& 
-\sum _{i,j=1}^2 \gamma _{ij} \left(\sigma_i^+ \sigma_j^- \, \rho + \rho \, \sigma_i^+ \sigma_j^- -2  \sigma_j^- \, \rho \, \sigma_i^+\right),
\end{align}
where $\sigma_i^+ = \ket{e_i} \bra{g_i}$ and $\sigma_i^- = \ket{g_i} \bra{e_i}$ are the usual raising and lowering operators for the $i$th atom, $2 \gamma_{ii} = 2 \Gamma$ is the Einstein A coefficient for spontaneous emission from the single atoms, and $\gamma_{ij} \; (i \neq j)$ and $g_{ij}$ are the collective parameters describing the damping rate and inter-atomic coupling that arises from the mutual influence of the atoms through the electromagnetic field [see App.~\ref{app:collective} for more detail].
By solving the master equation for the steady state, we determine a set of coupled equations of motion for the average values of the atomic operators and atomic correlations \cite{PRA52.636(1995)} that can be written in a vector form [see App.~\ref{append:dynamics} for more detail]:
\begin{align}
    \dot{\mathbf{S} }(t) = \mathbf{M} \, \mathbf{S} (t)  + \mathbf{b},
\end{align}
where $\mathbf{M}$ is a $15 \times 15$ matrix, $\mathbf{S} (t)$ is a column vector,
\begin{align}
    \mathbf{S} &\equiv \left( \tilde{\sigma}^+_1, \; \tilde{\sigma}^-_1 , \; \tilde{\sigma}^+_2, \; \tilde{\sigma}^-_2 , \; \tilde{\sigma}^+_1 \tilde{\sigma}^-_1 , \;
\tilde{\sigma}^+_2 \tilde{\sigma}^-_2 ,\;
\tilde{\sigma}^+_1  \tilde{\sigma}^-_2 ,\;
\tilde{\sigma}^+_2 \tilde{\sigma}^-_1 , \right. \nonumber \\
&\quad \; \tilde{\sigma}^+_1  \tilde{\sigma}^+_2 , \;
\tilde{\sigma}^-_1 \tilde{\sigma}^-_2 ,\;
\tilde{\sigma}^+_1  \tilde{\sigma}^-_1 \tilde{\sigma}_2^-  ,\; \tilde{\sigma}^+_1  \tilde{\sigma}^+_2  \tilde{\sigma}_1^- , \;
\tilde{\sigma}^+_2 \tilde{\sigma}^-_1  \tilde{\sigma}_2^-  ,  \nonumber \\
&\quad \left. \tilde{\sigma}^+_1  \tilde{\sigma}^+_2  \tilde{\sigma}_2^- ,\; \tilde{\sigma}^+_1  \tilde{\sigma}^+_2  \tilde{\sigma}^-_1 \tilde{\sigma}_2^- \right)^\text{T},
\end{align}
with
\begin{align}
\tilde{\sigma}^\pm_i (t) = \sigma^\pm_i \exp \pc{\mp \mathrm{i} \omega_L t},
\end{align}
and
\begin{align}
   \mathbf{b} \! \equiv \! \left(
- \frac{\text{i} \Omega_1^* }{2} ,\frac{\text{i} \Omega_1}{2}  ,
- \frac{\text{i} \Omega_2^* }{2} , \frac{\text{i} \Omega_2}{2}  ,
 0 ,  0 , 0 , 0 , 0 , 0 , 0 , 0 , 0 , 0 , 0 \right)^\text{T}
\end{align}
where the T indicates ``transpose''.

\subsection{Photon statistics $g^{(2)} (t)$}
The radiative properties of the atomic system can be characterized using the second-order correlation function $g^{(2)}(\tau)$. Study of it reveals directly if a field is quantum or classical. Coherent light has Poissonian statistics and is characterized by $g^{(2)} (\tau) = 1$. In a thermal (bunched) case, the magnitude of fluctuations is greater than that for a coherent state; $g^{(2)} (\tau)  > 1$, and the emitted light has super-Poissonian statistics. In opposition, for $g^{(2)} (\tau)  < 1$ or anti-bunched light, the photons have sub-Poissonian statistics. The most fundamental light source is an emitter of single photons --- that is, an emitting field that emits a single photon at a time --- corresponding to $g^{(2)} (0) = 0$.
To determine the photon statistics at position $\mathbf{R}$, we estimate the electric field operator $\hat{\mathbf{E}}^{(+)} \pc{\mathbf{R},t}
=\hat{\mathbf{E}}^{(+)}_\text{f} \pc{\mathbf{R},t}+\hat{\mathbf{E}}^{(+)}_\text{sf} \pc{\mathbf{R},t}$, where $\hat{\mathbf{E}}^{(+)}_\text{f} \pc{\mathbf{R},t}$ is the incident field and $\hat{\mathbf{E}}^{(+)}_\text{sf} \pc{\mathbf{R},t}$ is the radiation field of the atomic dipole moment, known as the source-field term. For an atom at position $\mathbf{r}_i$, the source-field term in the far field, $k \md{\mathbf{R}-\mathbf{r}_i} \gg 1$ for all $i$ [see Fig.~\ref{fig:ExpSetup}], is given by
\begin{align}
\hat{\mathbf{E}}^{(+)}_\text{sf} \pc{\mathbf{R},t} &= - \frac{k^2 \pc{\mathbf{d}_{eg} \times \hat{\mathbf{R}}}\times \hat{\mathbf{R}}}{4 \pi \varepsilon_0 r} 
\sum_{i=1}^2 e^{-\text{i} k \hat{\mathbf{R}} \cdot \mathbf{r}_i} \sigma_i^- \pc{t-\frac{R}{c}} .
\end{align}
Therefore, the source-field expression relates the scattered electric field to the properties of the atomic system. Detailed calculations show that the auto-correlation function for two laser driven two-level atoms is given by \cite{PhysRep372.369(2002)}
\begin{align}
g^{(2)} \pc{\mathbf{R}_1,\mathbf{R}_2; t, t+\tau} = \frac{G^{(2)} \pc{\mathbf{R}_1,\mathbf{R}_2; t, t+\tau}}{G^{(1)} \pc{\mathbf{R}_1,t} G^{(1)} \pc{\mathbf{R}_2,t+\tau} } \label{eq:g2},
\end{align}
which is a measure of the probability of detecting a photon at time $t + \tau$ at position $\mathbf{R}_2$, assuming the detection of a previous emission having occurred at time $t$ at position $\mathbf{R}_1$.
In experiments, the photon statistics are usually characterized by the steady-state second-order correlation function $g^{(2)}_\text{ss} (t \to \infty)$.

As in our study the correlation function is measured at a point in the far-field zone then we consider $ \hat{\mathbf{R}} =\hat{\mathbf{R}}_1 =\hat{\mathbf{R}}_2 $, and $\hat{\mathbf{R}} = \mathbf{R}/ R$ becomes the direction of observation [see Fig.~\ref{fig:ExpSetup}]. The first and second-order correlation functions are defined as
\begin{align}
   \frac{ G^{(1)} (\mathbf{R}, t)}{f \pc{\mathbf{R}}} &=   \sum_{i, j=1}^2 \av{\sigma^+_i (t) \, \sigma^-_j (t)}  \exp \pc{\text{i} k \hat{\mathbf{R}}\cdot \mathbf{r}_{ij}}
\end{align}
and
\begin{align}
   \frac{ G^{(2)} (\mathbf{R}; 0, t) }{f^2 \pc{\mathbf{R}} }&=  \sum_{i, j, k,l=1}^2 \av{\sigma^+_i (0) \, \sigma^+_k (t) \, \sigma^-_l (t) \, \sigma^-_j (0)} \nonumber \\
    &\quad \times \exp \pr{\text{i} k \pc{\hat{\mathbf{R}}\cdot (\mathbf{r}_{ij} + \mathbf{r}_{kl})} }
\end{align}
where $f(\mathbf{R})$ is a constant which depends on the geometry of the system, such as the angle between the observation direction and the atomic dipole moment. 
The first-order correlation function for two atoms reduces to
    \begin{align}
    \frac{ G^{(1)} (\mathbf{R}, t) }{ f \pc{\mathbf{R}} } &= \av{\sigma^+_1 (t) \, \sigma^-_1 (t)} + \av{\sigma^+_2 (t) \, \sigma^-_2 (t)} \nonumber \\
    &\quad + \av{\sigma^+_1 (t) \, \sigma^-_2 (t)} \exp \pc{\text{i} k \hat{\mathbf{R}} \cdot \mathbf{r}_{12}} \nonumber \\
    &\quad + \av{\sigma^+_2 (t) \, \sigma^-_1 (t)} \exp \pc{ - \text{i} k \hat{\mathbf{R}} \cdot \mathbf{r}_{12}}.
    \label{eq:G1}
    \end{align}
The two-time second-order correlation function $G^{(2)} \pc{\mathbf{R}; 0, t}$ yields terms of different forms. Following the arguments in Ref.~\cite{JPhysB49.025301(2016)}: 
(i) $\av{\sigma^+_j (0) \, \sigma^+_j (t) \, \sigma^-_j (t) \, \sigma^-_j (0)}$, i.e., terms with the same index, which correspond to the single-atom contributions; 
(ii) for $i \neq j$, $\av{\sigma^+_j (0) \, \sigma^+_i (t) \, \sigma^-_i (t) \, \sigma^-_j (0)} = \av{\sigma^+_j (0) \, \sigma^-_j (0) \, \sigma^+_i (t) \, \sigma^-_i (t) } $ and (iii) $\av{\sigma^+_i (0) \, \sigma^+_j (t) \, \sigma^-_i (t) \, \sigma^-_j (0)}$ are terms that involve two different atoms and that can be solved following an appropriate form of the quantum regression theorem \cite{BookCarmichael}; 
(iv) terms with $i \neq j$ in the form $\av{\sigma^+_i (0) \, \sigma^+_i (t) \, \sigma^-_j (t) \, \sigma^-_j (0)}$ are related to the anomalous correlation, which, for a thermal cloud, vanish on time averaging \cite{JPhysB49.025301(2016)}; and, finally, (v) $\av{\sigma^+_i (0) \, \sigma^+_j (t) \, \sigma^-_k (t) \, \sigma^-_i (0)}$, and the various permutations thereof,  drop out on time averaging due to the their random phases \cite{JPhysB49.025301(2016)}. Combined, this leads to
    \begin{align}
    \frac{G^{(2)} (\mathbf{R};0, t) }{f^2 \pc{\mathbf{R}}}
    &=  \av{\sigma^+_1 (0) \, \sigma^+_1 (t) \, \sigma^-_1 (t) \, \sigma^-_1 (0)} \nonumber \\
    &\quad +  \av{\sigma^+_2 (0) \, \sigma^+_2 (t) \, \sigma^-_2 (t) \, \sigma^-_2 (0)} \nonumber \\
    &\quad + \av{\sigma^+_1 (0) \, \sigma^+_2 (t) \, \sigma^-_1 (t) \, \sigma^-_2 (0)} \nonumber \\
    &\quad +  \av{\sigma^+_2 (0) \, \sigma^+_1 (t) \, \sigma^-_2 (t) \, \sigma^-_1 (0)} \nonumber \\
    &\quad +  \av{\sigma^+_2 (0) \, \sigma^+_1 (t) \, \sigma^-_1 (t) \, \sigma^-_2 (0) } \nonumber \\
    &\quad+ \av{\sigma^+_1 (0) \, \sigma^+_2 (t) \, \sigma^-_2 (t) \, \sigma^-_1 (0)}.
    \end{align}
In Ref.~\cite{JPhysB50.014004(2017)}, the exact dynamics of a disordered three-dimensional (3D) gas of up to $N=5$ atoms was solved. However, in our case, as $N \gg 1$, solving the exact dynamics of the system would become rather complex, as the dimensions of the density matrix $\rho$ grow with $2^N \times 2^N$. To overcome this limitation, we will solve the dynamics for two random pairs and average the photon statistics of multiple different random pairs. In the limit of many atoms, the first two terms in the equation, which represent the single-atom contributions, become relatively unimportant compared to the two-atoms' contributions. However, if we calculate the photon statistics by averaging over multiple pairs, this would, by default, lead to $\pc{N-1}$ counting of the single-atom contributions, which disagrees with our previous statement. To correct this over-counting, as we consider a large number of atoms, we neglect in our calculations the single-atom contributions when we calculate $G^{(2)} (\mathbf{R};0, t) $, which therefore simplifies to
    \begin{align}
    \frac{G^{(2)} (\mathbf{R};0, t) }{f^2 \pc{\mathbf{R}}}
    &=  \av{\sigma^+_1 (0) \, \sigma^+_2 (t) \, \sigma^-_1 (t) \, \sigma^-_2 (0)} \nonumber \\
    &\quad+  \av{\sigma^+_2 (0) \, \sigma^+_1 (t) \, \sigma^-_2 (t) \, \sigma^-_1 (0)} \nonumber \\
    &\quad+  \av{\sigma^+_2 (0) \, \sigma^+_1 (t) \, \sigma^-_1 (t) \, \sigma^-_2 (0) } \nonumber \\
    &\quad + \av{\sigma^+_1 (0) \, \sigma^+_2 (t) \, \sigma^-_2 (t) \, \sigma^-_1 (0)}.
    \label{eq:G2final}
    \end{align}
Finally, we must also consider the different terms contributing to the denominator in Eq.~\eqref{eq:g2},
\begin{widetext}
    \begin{align}
    \frac{ G^{(1)} (\mathbf{R}, 0)\,  G^{(1)} (\mathbf{R}, t) }{ f^2 \pc{\mathbf{R}} } &=  \pr{ \av{\sigma^+_1 (0) \, \sigma^-_1 (0)} + \av{\sigma^+_2 (0) \, \sigma^-_2 (0)} + \av{\sigma^+_1 (0) \, \sigma^-_2 (0)} \exp \pc{\text{i} k \hat{\mathbf{R}} \cdot \mathbf{r}_{12} } + \av{\sigma^+_2 (0) \, \sigma^-_1 (0)} \exp \pc{ - \text{i} k \hat{\mathbf{R}} \cdot \mathbf{r}_{12}}} \nonumber \\
    &\quad  \times \pr{ \av{\sigma^+_1 (t) \, \sigma^-_1 (t)} + \av{\sigma^+_2 (t) \, \sigma^-_2 (t)} + \av{\sigma^+_1 (t) \, \sigma^-_2 (t)} \exp \pc{\text{i} k \hat{\mathbf{R}} \cdot \mathbf{r}_{12}} + \av{\sigma^+_2 (t) \, \sigma^-_1 (t)} \exp \pc{ - \text{i} k \hat{\mathbf{R}} \cdot \mathbf{r}_{12}}}.
    \end{align}
    For a thermal vapor, due to averaging over the essentially random $\exp \pc{ \pm \text{i} k \hat{\mathbf{R}} \cdot \mathbf{r}_{12}}$ and $\exp \pc{ \pm 2 \text{i} k \hat{\mathbf{R}} \cdot \mathbf{r}_{12}}$ phase terms, this can be reduced to
    \begin{align}
    \frac{ G^{(1)} (\mathbf{R}, 0) \, G^{(1)} (\mathbf{R}, t) }{ f^2 \pc{\mathbf{R}} } &=  \av{\sigma^+_1 (0) \, \sigma^-_1 (0)} \av{\sigma^+_1 (t) \, \sigma^-_1 (t)} +  \av{\sigma^+_2 (0) \, \sigma^-_2 (0)}  \av{\sigma^+_2 (t) \, \sigma^-_2 (t)} + \av{\sigma^+_1 (0) \, \sigma^-_1 (0)} \av{\sigma^+_2 (t) \, \sigma^-_2 (t)} \nonumber \\
    &\quad  +  \av{\sigma^+_2 (0) \, \sigma^-_2 (0)} \av{\sigma^+_1 (t) \, \sigma^-_1 (t)}  + \av{\sigma^+_1 (0) \, \sigma^-_2 (0)} \av{\sigma^+_2 (t) \, \sigma^-_1 (t)} + \av{\sigma^+_2 (0) \, \sigma^-_1 (0)} \av{\sigma^+_1 (t) \, \sigma^-_2 (t)}.
    \end{align}
    Moreover, in the limit of many atoms, we similarly neglect in our calculation the single-atom contributions from our pair averaging, considering only the terms
    \begin{align}
    \frac{ G^{(1)} (\mathbf{R}, 0) \, G^{(1)} (\mathbf{R}, t) }{ f^2 \pc{\mathbf{R}} } &=  \av{\sigma^+_1 (0)\,  \sigma^-_1 (0)} \av{\sigma^+_2 (t) \, \sigma^-_2 (t)}  +  \av{\sigma^+_2 (0) \,  \sigma^-_2 (0)} \av{\sigma^+_1 (t) \, \sigma^-_1 (t)}  + \av{\sigma^+_1 (0) \, \sigma^-_2 (0)} \av{\sigma^+_2 (t) \, \sigma^-_1 (t)} \nonumber \\
    &\quad 
    + \av{\sigma^+_2 (0) \, \sigma^-_1 (0)} \av{\sigma^+_1 (t) \, \sigma^-_2 (t)}.
    \end{align}

Thus, we have the first- and second-order correlation functions expressed in terms of the atomic operator correlation functions. We can now directly apply our steady-state solutions of the atomic operators to calculate the photon statistics of the system,
\begin{align}
    \mathbf{S}_\text{ss} \equiv \mathbf{S} \pc{t \to \infty} = - \mathbf{M}^{-1} \mathbf{b}.
\end{align}
However, the solution to the master equation only yields single time averages; to find the equation of motion for multi-time vectors, we must take the single time equation of motion, multiplying on the left-hand side by $\sigma^+_i (0)$ and on the right by $\sigma^-_j (0)$ \cite{BookCarmichael}:  
    \begin{align}
        \frac{d }{dt} \av{\sigma^+_i (0) \, \mathbf{S} (t) \, \sigma^-_j (0)} &= \mathbf{M}\av{\sigma^+_i (0) \, \mathbf{S} (t) \, \sigma^-_j (0)}  + \av{\sigma^+_i (0) \, \sigma^-_j (0)}\mathbf{b}, \nonumber\\
        &= \mathbf{M} \pr{ \av{\sigma^+_i (0) \,  \mathbf{S} (t) \, \sigma^-_j (0)}  + \av{\sigma^+_i (0) \, \sigma^-_j (0)} \mathbf{M}^{-1} \mathbf{b}}.
    \end{align}
    The formal solution to this equation is given by 
    \begin{align}
        \av{\sigma^+_i (0) \, \mathbf{S} (t) \, \sigma^-_j (0)} &=
        - \av{\sigma^+_i \sigma^-_j}_\text{ss} \mathbf{M}^{-1} \mathbf{b} + \exp (\mathbf{M} t) 
        \pr{\av{\sigma^+_i \, \mathbf{S} \, \sigma^-_j }_\text{ss}   + \av{\sigma^+_i \sigma^-_j}_\text{ss} \mathbf{M}^{-1} \mathbf{b} }, \nonumber \\
        &= \exp (\mathbf{M} t) \av{\sigma^+_i \, \mathbf{S} \, \sigma^-_j }_\text{ss} + \av{\sigma^+_i \sigma^-_j}_\text{ss} 
        \pr{ \exp (\mathbf{M} t) - 1} \mathbf{M}^{-1} \mathbf{b} .
    \end{align}
Inserting this solution back into the expression determined for $G^{(2)} (\mathbf{R};0, t)$ in eq.~\eqref{eq:G2final}, confirms that $G^{(2)} (\mathbf{R};0, t=0) = G^{(2)}_\text{ss} (\mathbf{R}) $ [see App.~\ref{App:G2} for more detail].
\end{widetext}

Please note that, in this study, $g^{(2)} (\tau)$ accounts for all the emitted photons regardless of their frequencies. In our envisaged experiment, the photodetection setup is not considered to discriminate between the different-frequency photons over the relevant frequency range.
While it is true that in a thermal vapor, the emitted photons will in general be subject to Doppler shifts, broadening their frequency spectrum, the spectral response typically has a Doppler width of order 1~GHz, whereas photodetectors generally have a much larger bandwidth (commonly spanning the visible and near-infrared regime). While filters are often used experimentally to block stray light, these generally have bandwidths of order 10~nm, relative to which the Doppler width can still be considered insignificant. We note that the Doppler shift will also have a directional dependence, in that such an effect should be minimal when considering scattering in the forward direction. If, as is considered to be the case here, we do not discriminate between photons of different frequencies, averaging over the random orientations of the dipoles means that we do not observe any directional dependence, however (in Fig.\ref{fig:ExpSetup} the photodetector is oriented perpendicular to the propagation direction of the driving laser beam, for example).
We also note that having a very narrow band photon source is not always an advantage; this for example requires the photons to be ``long'' in the time domain. Nevertheless, if it were desirable to study the statistics of photons within a very narrow range of frequencies, this could be done by setting a narrow filter about a frequency within the spectrum of the source and re-calculating the photon statistics following Refs.~\cite{JPhysB16.2677(1983), JPhysB20.4915(1987)}.

\section{Numerical Results} 
The purpose of our studies is to model the first- and second-order correlation functions of the electromagnetic field at a point $\mathbf{R}$ in the far-field zone of the radiation emitted by the atomic system [see Fig.~\ref{fig:ExpSetup}]. In the case of the rubidium D2 line of interest in this paper, the atomic transition wavelength is $\lambda = 780$~nm, and the decay rate is $\Gamma = 2 \pi \times 6$~MHz. We give our numerical results in units of decay rate $\Gamma$ and wavelength $\lambda$, however, making them in principle more general. 
One of the advantages of working with alkali-metal vapors is the broad control over the number density obtainable by tuning the cell's temperature \cite{JKeaveneyThesis}.
When dealing with thermal vapors, the atoms are not static, as they collide with each other and with the surface of the cell enclosing the vapor. When an atom hits a surface, two things can happen: the atom bounces back from the surface elastically or sticks to the surface for a certain time before flying away. Atom-surface interactions often cannot be easily understood or controlled, and influence the atomic adsorption/desorption dynamics. These in general directly impact the atomic density and its relation to temperature. For simplicity, however, we neglect explicit consideration of such effects in this paper.

The expected photon statistics of the thermal vapor are calculated via a Monte Carlo simulation. Knowing the temperature of the vapor, we can compute the atomic number density, $N$. This is determined from the vapor pressure $p$, i.e., the gaseous phase's pressure in equilibrium with a solid or liquid bulk of the same material \cite{RubidiumData}, and the  temperature $T$. The vapor pressure is given by
\begin{align}
\log_{10} p = 2.881 + 4.312 - \frac{4040}{T},
\end{align}
and the atomic number density by
\begin{align*}
N = \frac{133.323 p}{k_\text{B} T},
\end{align*}
for the liquid phase of rubidium where, $k_\text{B}$ is the  Boltzmann constant. For a random 3D distribution of atoms, the spacing between the atoms $r$ has the distribution \cite{RevModPhys15.1(1943)}
\begin{align}
W_\text{3D} (r) = 4 \pi N r^2 e^{- 4 \pi / 3 N r^3}.
\end{align}
The average distance between the atoms can then be found from the atomic number density to be
\begin{align}
r_\text{av} = \int_0^\infty r W(r) dr \approx \frac{5}{9} N^{-1/3}.
\end{align}
We can now randomly place two atoms inside a cubic box, $\pc{0,\, L}\times \pc{-L/2,\,L/2} \times \pc{-L/2,\,L/2}$, the size of which ensures the average spacing. Moreover, we set a minimum distance at which the two atoms can be placed at, i.e., if an atom is at the center of a sphere, there will be at most one other atom within a distance that depends on the interaction strength and the mean relative velocity of the atoms. For our simulations, we set this minimum possible distance between the atoms to be $0.01 \lambda$, where, beyond that, other perturbations to the model should be considered.

The atomic motion will also introduce Doppler broadening of absorption lines from the natural linewidth --- to account for such effects in our numerical simulations, a random velocity is attributed to each particle following a probability based on the Boltzmann distribution, which is equivalent to randomly attributing a laser detuning to each atom \cite{PRL112.113603(2014),PRA96.033835(2017)}. Thus, to mimic the effects of temperature, we give each atom an individual effective detuning. The average detuning is set $\Delta_\text{av} / \Gamma$; in our numerics $\Delta_i$ can then have any value on the interval $\Delta_i = \pr{\Delta_\text{av}-5 \, \Gamma, \, \Delta_\text{av} +5 \, \Gamma}$ according to probabilities based on the Boltzmann distribution. However, although we include the Doppler effect on the excitation, in our model we neglect atomic motion during the emission. Having the detector placed at infinity, the movement of the atom during emission can be safely neglected; if the atom moves outside the detector spot, it will not contribute to the photon statistic measurements.
Having set the vapor's temperature, we compute the atomic density and, thus, the average distance between two atoms. We apply Eq.~\eqref{eq:g2} for a pair of atoms randomly placed inside a cubic box that ensures the defined average spacing, and fixing the driving direction along the $x$-direction [see Fig.~\ref{fig:ExpSetup}]. This process is repeated and averaged over 1500 different realizations. 
\begin{figure}[t]
\centering
\includegraphics[width=\linewidth, keepaspectratio]{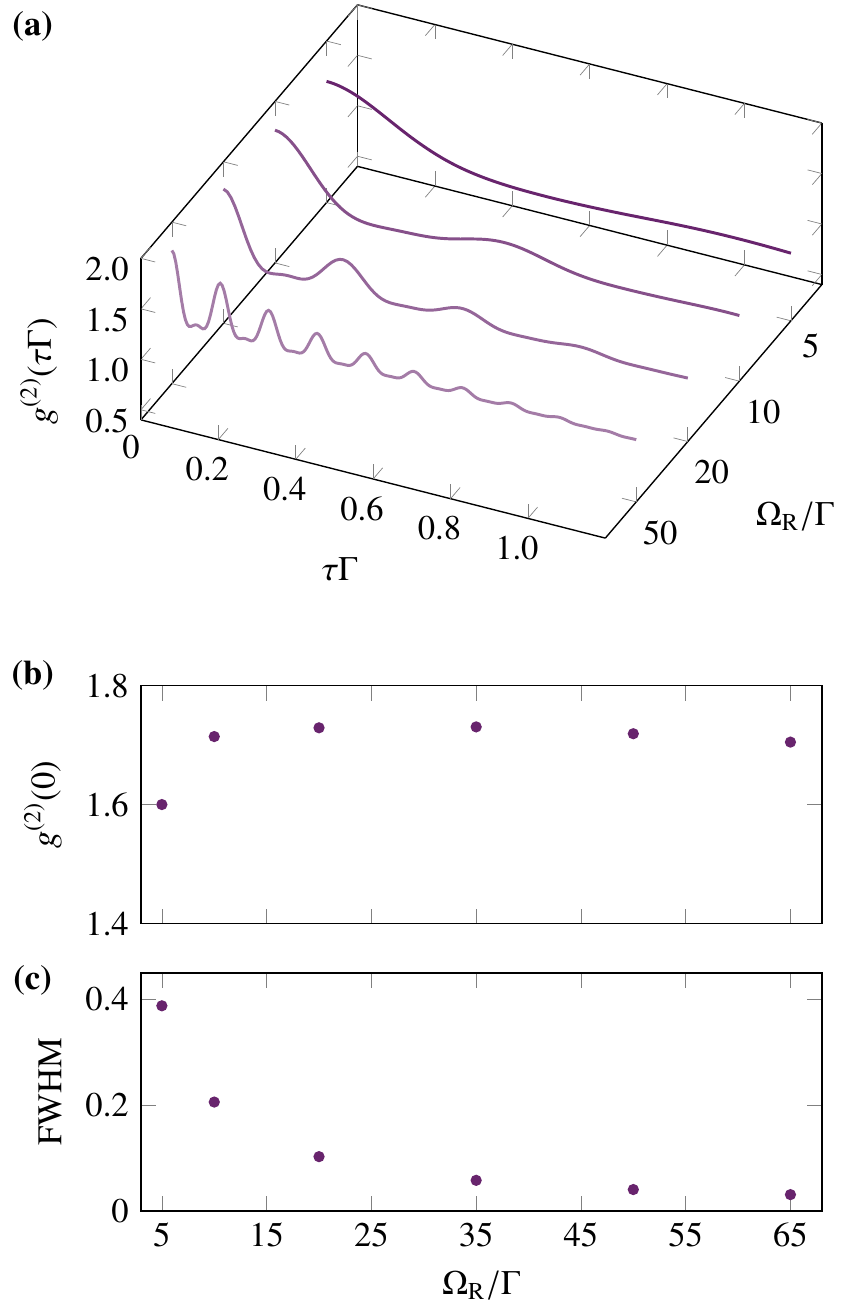}
\caption{(Color online) Photon statistics for multiple atomic pairs in a running-wave laser field with different driving frequencies, where we have fixed the average values of detuning for both atoms at $\Delta_\text{av}/\Gamma$ equal to zero and $T=380$~K (or $\av{r_{12}/ \lambda } \sim 0.35$). 
(a) $g^{(2)} (\tau \Gamma)$ as a function of time for different driving Rabi frequencies $\Omega_\text{R} / \Gamma = $5, 10, 15, 50. (b)  $g^{(2)} (\tau \Gamma = 0)$ and (c) first peak FWHM of the $g^{(2)}(\tau \Gamma)$ as functions of different driving Rabi frequencies. For each data point, we run the simulation over 1500 pairs. Our units are chosen such that the quantities are scaled by $\lambda$ or $\Gamma$. \label{Fig:Driving}}
\end{figure}

\begin{figure}[t]
\centering
\includegraphics[width=\linewidth, keepaspectratio]{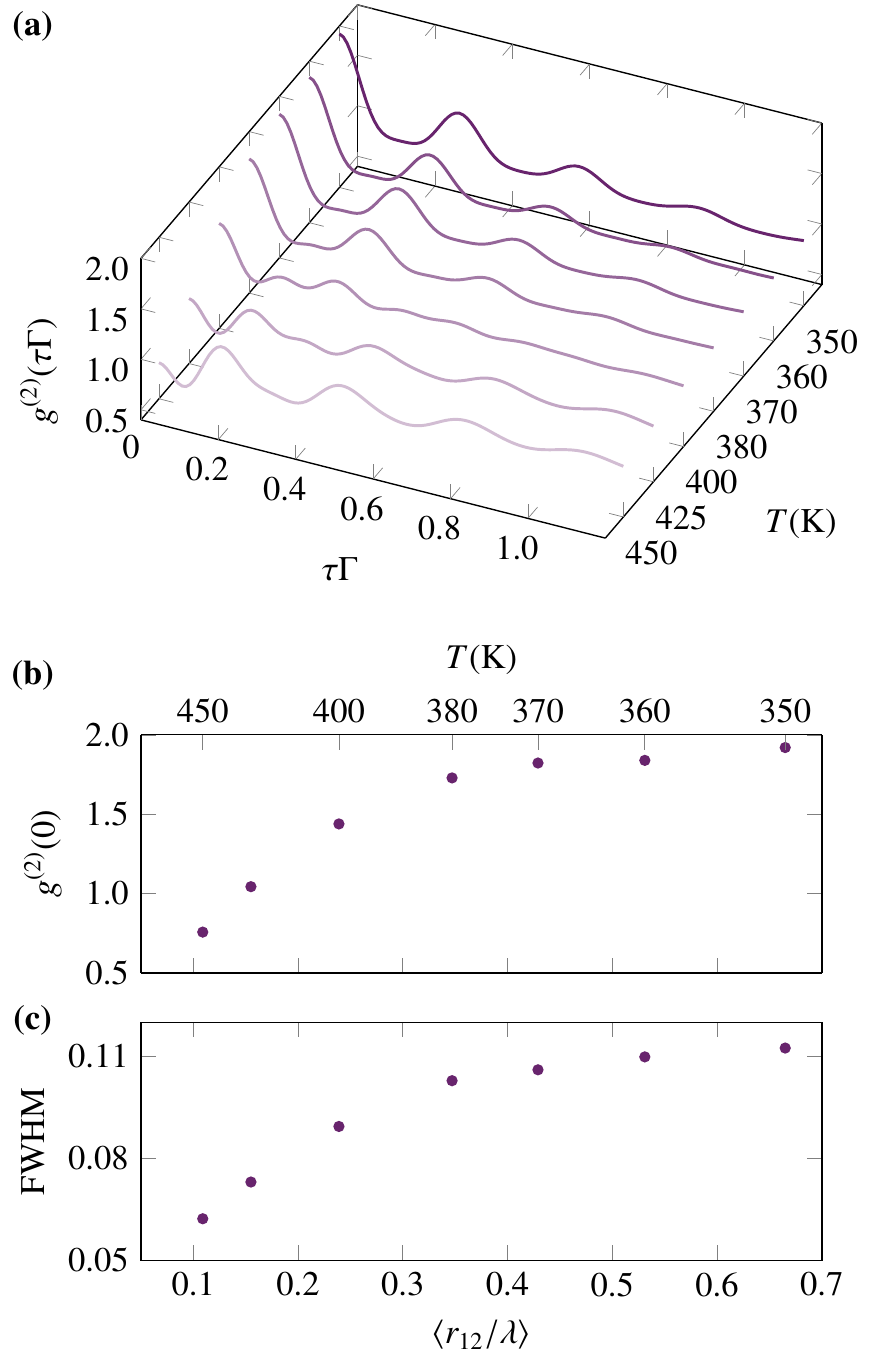}
\caption{(Color online) Photon statistics for multiple atomic pairs in a running-wave laser field with $\Omega_\text{R} / \Gamma = 20$, where we have fixed the average values of detuning for both atoms at $\Delta_\text{av}/\Gamma$ equal to zero, for different temperatures. (a) $g^{(2)} (\tau \Gamma)$ as a function of time for different temperatures $T = 450$ K, 425 K, 400 K, 380 K, 370 K, 360 K, and 350 K. (b)  $g^{(2)} (\tau \Gamma = 0)$ and (c) first peak FWHM of the $g^{(2)}(\tau \Gamma)$ as functions of different averaged inter-atomic distances (or, equivalently, temperature) for multiple atomic pairs . For each data point, we run the simulation over 1500 pairs. Our units are chosen such that the quantities are scaled by $\lambda$ or $\Gamma$. \label{Fig:Temperatures}}
\end{figure}

\begin{figure}[t]
\centering
\includegraphics[width=\linewidth, keepaspectratio]{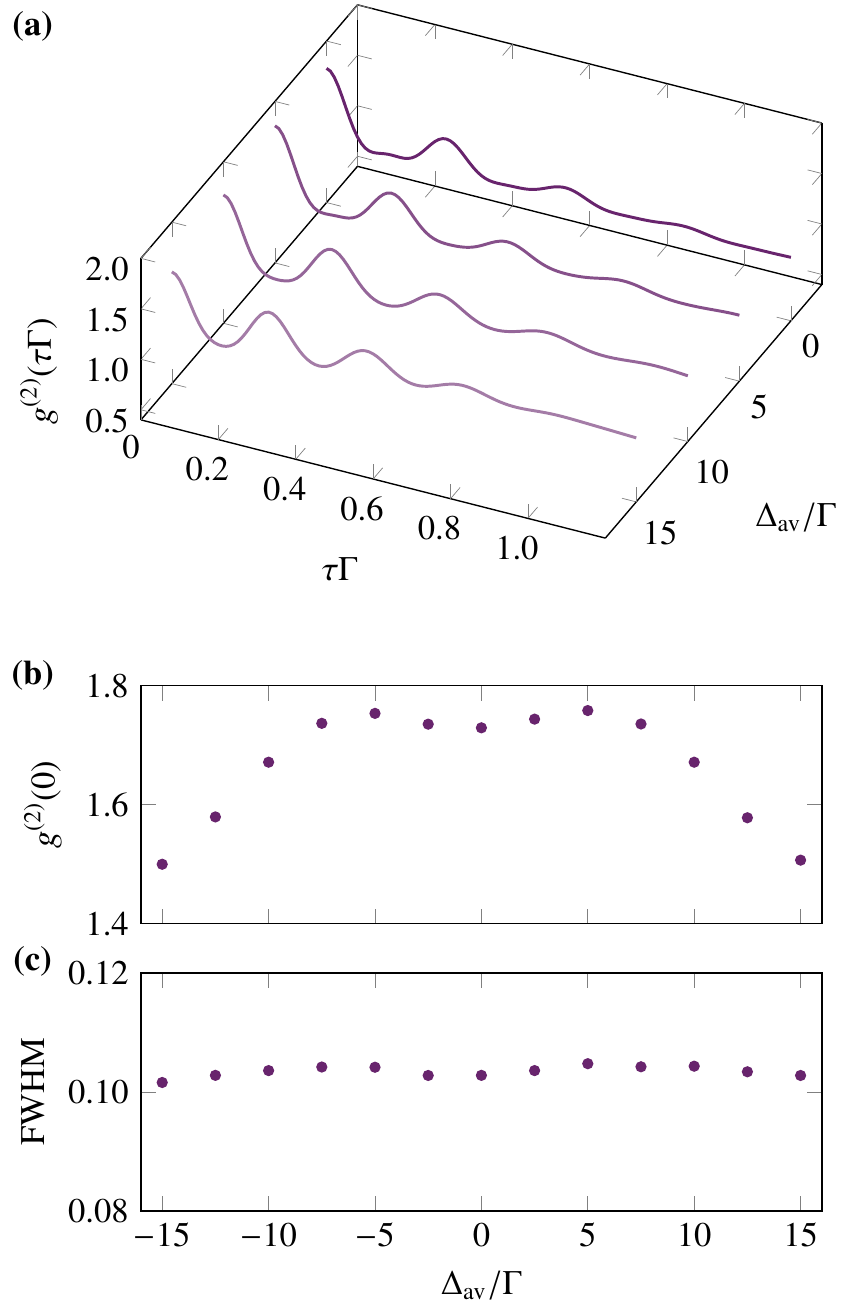}
\caption{(Color online) Photon statistics for multiple atomic pairs in a running-wave laser field with $\Omega_\text{R} / \Gamma = 20$, where we have fixed the average values of detuning for both atoms at $\Delta_\text{av}/\Gamma$ and $T=380$~K (or $\av{r_{12}/ \lambda } \sim 0.35$). (a) $g^{(2)} (\tau \Gamma)$ as a function of time for different averaged detunings $\Delta_\text{av}/\Gamma=0,\, 5,\, 10$ and 15, similar results obtained for negative values of detuning. (b)  $g^{(2)} (\tau \Gamma = 0)$ and (c) first peak FWHM of the $g^{(2)}(\tau \Gamma)$ as functions of different averaged detunings. For each data point, we run the simulation over 1500 pairs. Our units are chosen such that the quantities are scaled by $\lambda$ or $\Gamma$. \label{Fig:Detuning}}
\end{figure}

\begin{figure}[t]
\centering
\includegraphics[width=\linewidth, keepaspectratio]{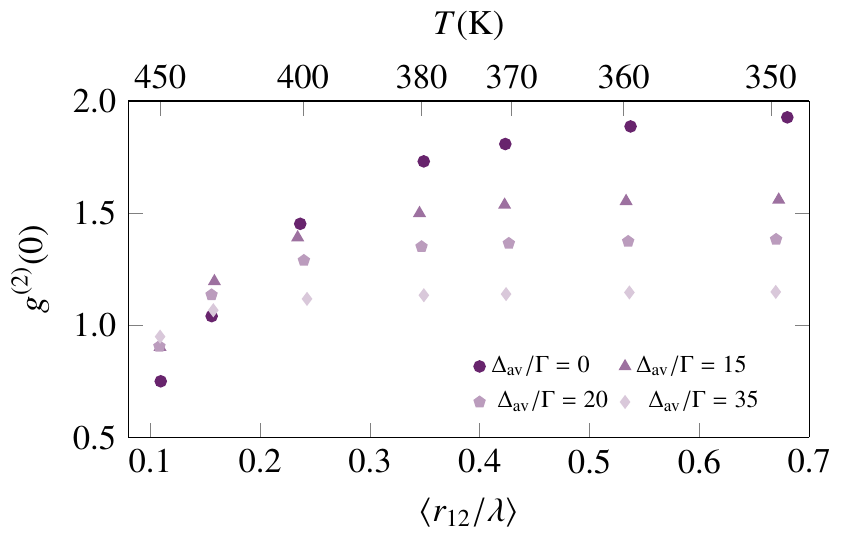}
\caption{(Color online) Photon statistics for multiple atomic pairs in a running-wave laser field with $\Omega_\text{R} / \Gamma = 20$, where we have fixed the average values of detuning for both atoms at $\Delta_\text{av}/\Gamma = 0$ (circles), 15 (triangles), 20 (diamonds), and 35 (pentagons) for different atomic densities. 
For each numerical data point, we run the simulation over 1500 pairs.
\label{Fig:DensityDetuning}}
\end{figure}

The presented model predicts varied behavior for the photon statistics for a single pair of atoms under different experimental conditions. We numerically simulated $g^{(2)} (\tau)$ for different driving [see Fig.~\ref{Fig:Driving}], temperatures [see Fig.~\ref{Fig:Temperatures}], and detuning [see Fig.~\ref{Fig:Detuning}]. From the time evolution of the photon statistics, it is possible to retrieve $g^{(2)} (\tau = 0)$ and estimate the full width at half maximum (FWHM) of the first oscillation. 
Periodic oscillations are predicted in the photon statistics of a thermal gas for all our numerical simulations. These are related to the coherent Rabi oscillations that the two-level atoms undergo. In our model, the atoms are driven at the same frequency and are more likely to emit at the same regular interval with a periodicity related to the inverse of the Rabi frequency. In an experimental configuration where the spatial profile of the driving laser is relevant, these oscillations will be difficult to observe, and the multi-level nature of real atoms may add further complications.
We observe that the photon statistics decay with time as decoherence sets in due to spontaneous emission \cite{JPhysB49.025301(2016)}, and the value of $g^{(2)} (\tau)$ approaches unity for $\tau \to \infty$. From Figs.~\ref{Fig:Driving}(b) and \ref{Fig:Driving}(c), we can observe that the driving frequency does not significantly affect the value at $\tau=0$, but, as explained above, the FWHM decreases as we increase the driving frequency. As the laser field becomes stronger, it dominates over the different dynamics of the system, and the value of $g^{(2)} (0)$ saturates. Although we could intuitively expect to detect an output light that is characteristic of the laser field, with $g^{(2)} (0) = 1$, as the intensity of the driving increases, the scattering of the laser light when passing the (random) thermal vapor will prevent this from happening. For temperatures where the dipole-dipole interaction can be neglected, as we increase the driving field, the value of $g^{(2)} (0) \to 2$, as is typical for a thermal light source. As we increase the temperature and the dipole-dipole interactions become more and more important, this value decreases.

We now discuss how the photon statistics of a thermal vapor vary with temperature or, equivalently, inter-atomic distance. In Figs.~\ref{Fig:Temperatures}(b) and \ref{Fig:Temperatures}(c), we show the results for our model of the photon statistics $g^{(2)}(0)$ of a thermal vapor. As the average inter-atomic distance increases, the $g^{(2)}(0)$ function reaches a plateau; the plateau value differs depending on the different averaged values of detuning, see Fig.~\ref{Fig:DensityDetuning}. For two two-level atoms, when the laser frequency is tuned to the atomic resonance, i.e., $\Delta_\text{av} / \Gamma= 0$, for small distances between the atoms, the dipole-dipole interaction becomes so large that we can never observe an anti-bunching effect. This will occur for weak and strong driving \cite{PRA58.4133(1998)}. However, in the case of a thermal vapor, the increase of the dipole-dipole interactions leads to a different response of the photon statistics. We observe that the strong light-induced interactions will increase anti-bunching for small atomic distances. This behavior is maintained even for  $\Delta_\text{av} / \Gamma \neq 0$.
However, at low densities, i.e. large inter-atomic distances and weaker dipole-dipole interactions, for $\Delta_\text{av} / \Gamma = 0$, we see the plateau approaching $g^{(2)} (0) \to 2$ more characteristic of a classic thermal source. The FWHM also decreases with the increase of the dipole-dipole interactions; by decreasing the atomic distance, we are also increasing the decay rate [see eq.~\eqref{eq:apgamma12} in App.~\ref{app:collective}], which leads to a reduction of the coherence time.
Finally, with Figs.~\ref{Fig:Detuning}(b) and \ref{Fig:Detuning}(c), we explore how the detuning affects the photon statistics of the thermal vapor. To do so, we define the temperatures, $T= 380$~K, which correspond to an average distance of $0.35 \lambda$. We can see that $g^{(2)} (\tau =0)$ is almost symmetric around zero detuning, and for larger values of detuning, the values $g^{(2)} (\tau =0)$ decrease. However, there is no significant  dependence of the FWHM on the detuning. In a system with two two-level atoms, the anti-bunching effect can be increased for finite values of detuning, in particular when the dipole-dipole interaction and detuning cancel each other \cite{PRA29.2004(1984)}. In this special situation, the driving only targets the symmetric state, and we observe that the strong light-induced interactions will increase anti-bunching for small atomic distances. This phenomenon occurs as the driving is done between the ground-state and the system's collective bright state, and the two-atom system behaves effectively as a single two-level system. This has been proven to be true even for arrays of atoms as long as we are able to drive the collective bright state of the ensemble \cite{PRL125.073601(2020)}. Although there will not be such a classic blockade effect in thermal vapors, our results show that an increase of anti-bunching can still be observed for finite values of detuning.

In Fig.~\ref{Fig:DensityDetuning}, we compare how the $g^{(2)} (0)$ varies with density for different values of average detuning. It is clear that it is possible to tune the value of the $g^{(2)} (0)$ by changing the detuning and temperature of the thermal vapor. Increasing the value of the detuning will lead to a flattening of the curve to values of $g^{(2)} (0) = 1$. Furthermore, for higher temperatures, it is theoretically possible, within this model, to achieve $g^{(2)}(0) < 1$, characteristic of sub-Poissonian statistics. 
However, at higher temperatures, i.e., smaller inter-atomic distances and stronger dipole-dipole interactions, we expect this model to predict that the photon statistics value of $g^{(2)}$ decreases much faster than we would observe in an experiment. In fact, in Ref.~\cite{WillThesis} where the author presents a study of the photon statistics of a thermal vapor confined in a nano-cell \cite{KWhittaker2015, KWhittakerThesis, Tom}, we observe exactly that. 
The confinement regimes of order $\lambda$ within these nano-cells means that the atoms can be closer to the surface, experiencing the typical atom-surface shift of hundreds of megahertz, corresponding roughly to $\Delta/\Gamma \sim 15$ \cite{KWhittakerThesis}. By comparing both results, the developed theoretical model agrees with the photon statistics data presented in Ref.~\cite{WillThesis}.
The expected discrepancy at relatively small distances could be due to effects beyond our theoretical models' limits, such as collisional and motional dephasing of the dipoles or three-body effects.  Experiments that explore the variation of detuning have not been performed yet. It would be interesting to explore if, as our predictions indicate, the increase in the detuning would lead to smaller values of $g^{(2)}(0)$ without the need to access higher densities. Nevertheless, by varying temperature, our theoretical predictions allow one to explore considerably different photon statistics regimes in a thermal vapor.

\section{Conclusions}
In conclusion, this paper presents a theory to calculate the photon statistics of an ensemble of atoms in thermal vapor cells. The qualitative agreement between the theory and the experiment demonstrates that the photon statistics can be well described by considering a simple model of pairwise interactions. The effects of atomic motion were taken into account in the expected Doppler shift. Our model allows for the exploration of different regimes of the $g^{(2)} (\tau)$ function in the thermal vapor. By varying the laser intensity, atomic density, and detuning, we moved between different photon statistics regimes.

Further experimental work will be needed to determine the necessary parameters to obtain $g^{(2)} (0) \to 0$. Moreover, current nano-cells' flexibility allows the production of arbitrary internal geometries and exploration of different dimensionality \cite{Tom}. A combination of these features may be a promising avenue to generate single-photon sources with thermal vapors.
Research into collective processes is an active area of study with many open questions. Our results provide useful guidelines for further developments, as more experiments and new technology arises. This knowledge can be exploited to design new experimental setups with the potential for new quantum technologies involving hot atomic vapors.

Additional data related to the findings reported in this paper are made available by the source in Ref.~\cite{dataDOI}. 

Acknowledgements: We would like to acknowledge the support from the UK Engineering and Physical Sciences Research Council Grant No. EP/R002061/1.
The authors are grateful to Ifan G. Hughes and W. J. Hamlyn for the fruitful discussions.

\appendix
\section{Ensemble of atomic emitters \label{app:collective}}

We begin by investigating the modifications to the individual single-atom decay rates and emerging collective energy shifts (we will follow closely Refs.~\cite{OstermannThesis,PhysRep372.369(2002)} and references therein). Assuming $N$ identical emitters with a transition frequency of $\omega_0$ and neglecting atomic motion or collisions, the Hamiltonian of the system is given by $\hH = \hH_\text{A} + \hH_\text{F} + \hH_\text{int}$, with
\begin{align}
\hH_\text{A} &= \hbar \omega_0 \sum_{i=1}^{N} \sigma^+_i \sigma^-_i,\\
\hH_\text{F} &= \hbar \sum_{\mathbf{k},\lambda} \omega_\mathbf{k} \ha^\dagger_{\mathbf{k},\lambda} \ha_{\mathbf{k},\lambda},\\
\hH_\text{int} &= \mathrm{i} \hbar \sum_{i=1}^N \sum_{\mathbf{k},\lambda} g_{\mathbf{k}, \lambda} \nonumber \\
&\quad \times
\pr{\ha_{\mathbf{k}, \lambda} \exp \pc{\mathrm{i} \mathbf{k} \cdot \mathbf{r}_i} - \text{H.c.} }
\pc{\sigma_i^+ + \sigma_i^-},
\end{align}
and with $g_{\mathbf{k}, \lambda} = \sqrt{\omega_k / (2 \varepsilon_0 V)} \mathbf{e}_{\mathbf{k}, \lambda} \cdot \mathbf{d}_{eg}$ where we assume an equal orientation and amplitude for the atomic transition dipoles $\mathbf{d}_i = \mathbf{d}_{eg} = \bra{e} \mathbf{d} \ket{g}$. 
The equation of motion for the field operators is given by
\begin{align}
\dot{\ha}_{\mathbf{k},\lambda} &= - \mathrm{i} \omega_k \ha_{\mathbf{k}, \lambda} - g_{\mathbf{k}, \lambda} \sum_{i=1}^N \exp \pc{- \mathrm{i} \mathbf{k}\cdot \mathbf{r}_i}
\pc{\sigma_i^+ + \sigma_i^-}.
\end{align}
This can be solved by means of a retarded Green function as
\begin{align}
\ha_{\mathbf{k},\lambda} (t) &= \ha_{\mathbf{k}, \lambda} \pc{t_0} \exp \pc{- \mathrm{i} \omega_k \pc{t-t_0}}  
\nonumber \\
&\quad - \int_{t_0}^t dt' \exp \pc{-\mathrm{i} \omega_k (t-t')} g_{\mathbf{k}, \lambda} \nonumber \\
&\quad \times
\sum_{i=1}^N \exp \pc{-\mathrm{i} \mathbf{k}\cdot \mathbf{r}_i} \pc{\sigma^+_i + \sigma^-_i}. 
\label{eq:sol_a}
\end{align}

Next, we will consider the equation of motion of any atomic operator $\hO$
\begin{align}
\dot{\hat{\mathrm{O}}} &= \mathrm{i} \omega_0 \sum_{i=1}^N \pr{\sigma^+_i \sigma^-_i, \hO}  \nonumber \\
&\quad - \sum_{\mathbf{k}, \lambda} \sum_{i=1}^N g_{\mathbf{k}, \lambda} 
\left(
\pr{\sigma^+_i +\sigma^-_i, \hO} \ha_{\mathbf{k}, \lambda} \exp \pc{\mathrm{i} \mathbf{k}\cdot \mathbf{r}_i} \right.
\nonumber \\
&\quad \left.
- \ha^\dagger_{\mathbf{k}, \lambda} \exp \pc{-\mathrm{i} \mathbf{k}\cdot \mathbf{r}_i} \pr{\sigma^+_i +\sigma^-_i, \hO} \right).
\end{align}
Inserting Eq.~\eqref{eq:sol_a} back into the equation of motion, we can replace $\textstyle\sum_{\mathbf{k},\lambda} \to V / (2 \pi)^3 \int d^3 k$ and abbreviating the contribution from the incident field by $E_\text{in} (t)$, one finds
\begin{align}
\dot{\hO} &= E_\text{in} (t) \nonumber\\
&\quad + \frac{\mathbf{d}_{eg}^2}{2 \varepsilon_0 (2 \pi c)^3} \sum_{i,j}
\int_{\Omega} d \Omega_\mathbf{k} \pc{1- \pc{e_{\mathbf{d}}\cdot e_\mathbf{k}}^2}  \nonumber \\
&\quad \times
\int_{t_0}^t d t' \int_0^\infty d \omega \,
\omega^3 \exp \pc{-\mathrm{i} \omega \pc{t-t' - e_\mathbf{k} \cdot \mathbf{r}_{ij}/c } } 
\nonumber \\
&\quad \times \pr{\sigma^x_i (t), \hO (t)} \sigma^x_j (t') \nonumber \\
&\quad - \frac{\mathbf{d}_{eg}^2}{2 \varepsilon_0 (2 \pi c)^3} \sum_{i,j}
\int_{\Omega} d \Omega_\mathbf{k} \pc{1- \pc{e_{\mathbf{d}}\cdot e_\mathbf{k}}^2} \nonumber \\
&\quad \times
\int_{t_0}^t d t' \int_0^\infty \!\! d \omega \, 
\omega^3 \exp \pc{\mathrm{i} \omega \pc{t-t' - e_\mathbf{k} \cdot \mathbf{r}_{ij}/c } }  \nonumber \\
&\quad \times
\sigma^x_j (t') \pr{\sigma^x_i (t), \hO (t)} ,
\end{align}
where we have used the relation $\sigma^x_i = \sigma^+_i +\sigma^-_i$, and the fact that $\mathbf{k} \perp e_{\mathbf{k},1} \perp e_{\mathbf{k},2}$, such that $\sum_\lambda \md{\mathbf{d}_{eg} \cdot e_{\mathbf{k},\lambda}}^2
= d_{eg}^2 \pc{1 - \pc{e_{\mathbf{d}} \cdot e_\mathbf{k}} }^2$,
where $e_{\mathbf{d}} = \mathbf{d}_{eg}/ d_{eg}$ and $e_\mathbf{k} = \mathbf{k}/k$.

Performing the Markov and rotating wave approximations, we integrate over the solid angle, finding
\begin{align}
\dot{\hat{\mathrm{O}}} &= E_\text{in} (t)  + \sum_{i,j} \pr{\sigma^x_i (t), \hO(t)} \nonumber \\
&\quad \times
\frac{\Gamma}{k_0^3} \int_0^\infty
\frac{d k}{2 \pi} k^3 \mathcal{F} \pc{k r_{ij}} \nonumber\\
&\quad\times
\left[
\pc{-\mathrm{i} \, \mathcal{P} \frac{1}{k+k_0} + \pi \delta \pc{k+k_0} } \sigma^+_j (t) \right. \nonumber \\
&\quad \left.  +\pc{-\mathrm{i} \, \mathcal{P} \frac{1}{k-k_0} + \pi \delta \pc{k-k_0}} \sigma^-_j(t) \right]
\nonumber \\
&\quad - \sum_{i,j}  \frac{\Gamma}{k_0^3} \int_0^\infty
\frac{d k}{2 \pi} k^3 \mathcal{F} \pc{k r_{ij}} \nonumber\\
&\quad \times \left[ \sigma^+_j (t) \pc{\mathrm{i}\, \mathcal{P} \frac{1}{k-k_0} + \pi \delta \pc{k-k_0} } \right. \nonumber \\
&\quad + \left. \sigma^-_j(t) \pc{\mathrm{i}\, \mathcal{P} \frac{1}{k+k_0} + \pi \delta \pc{k+k_0}} \right] \nonumber \\
&\quad \times \pr{\sigma^x_i (t), \hO(t)} ,
\end{align}
where we have defined
\begin{align}
\mathcal{F} \pc{k r_{ij}} = \frac{3}{2} \int_\Omega \frac{d \Omega_k}{4 \pi} \pc{1 - \pc{e_{\mathbf{d}} \cdot e_\mathbf{k}}^2} \exp \pc{\mathrm{i} \mathbf{k} \cdot \mathbf{r}}.
\end{align}

From this one can now describe the quantities corresponding to collective energy shifts and emission rates as
\begin{align}
\gamma_{ij} &= \Gamma \mathcal{F} \pc{k_0 r_{ij}}, \\
g^\pm_{ij} &= \frac{\Gamma}{k_0^3} \mathcal{P} \int_0^\infty \frac{d k }{2 \pi}
\frac{k^3 \mathcal{F} \pc{k r_{ij}} }{k \pm k_0}.
\end{align}
We now deduce the simplified equation of motion
\begin{align}
\hO &= E_\text{in} (t)  + \sum_{i,j} \pr{\sigma^x_i (t), \hO(t)}
\nonumber \\
&\quad \times
\px{\pc{-\mathrm{i} g^+_{ij}} \sigma^+_j (t) + \pc{-\mathrm{i} g^-_{ij} + \frac{\gamma_{ij}}{2}} \sigma^-_j(t) } \nonumber\\
&\quad - \sum_{i,j} 
\px{\pc{\mathrm{i} g^-_{ij} + \frac{\gamma_{ij}}{2}} \sigma^+_j (t) + \pc{\mathrm{i} g^+_{ij}} \sigma^-_j(t) }  \nonumber\\
&\quad \times
\pr{\sigma^x_i (t), \hO(t)}.
\end{align}
Dropping the rapidly oscillating terms, we arrive at
\begin{align}
\hO &= \mathrm{i} \sum_i 
\pr{\pc{\omega_0 - g^+_{ii}} \, \sigma^+_i (t) \, \sigma^-_i (t) - g^-_{ii}\, \sigma^-_i(t) \, \sigma_i^+ (t), \,\hO (t) }
\nonumber \\
&\quad
+ \mathrm{i} \sum_{i \neq j} \pr{g_{ij} \, \sigma^-_i (t) \, \sigma^+_j (t),\, \hO(t)} \nonumber \\
&\quad + \frac{1}{2} \sum_{i,j} \gamma_{ij} 
\left(
2 \sigma^+_i (t)\, \hO (t) \, \sigma^-_j (t) - \sigma^+_i (t) \, \sigma^-_j(t) \hO(t) \right. \nonumber \\
&\quad - \left. \hO(t) \, \sigma^+_i(t) \, \sigma^-_j(t) \right) ,
\end{align}
where we may easily recognize a quantum Langevin equation. Returning to the Schr\"odinger picture and writing the equation of motion for the density operator $\rho$ describing the mixed state of the system
\begin{align}
\dot{\rho} = \frac{\mathrm{i}}{\hbar} \pr{\rho, \hH} + \mathcal{L} \pr{\rho},
\end{align}
we find
\begin{align}
\hH &= \hbar \sum_i \omega_0 \sigma^+_i \sigma^-_i 
+ \frac{\hbar}{2} \sum_{i\neq j} g_{ij} \pc{ \sigma_i^- \sigma_j^+ +\text{H.c.}}, \\
\mathcal{L} \pr{\rho} &= \sum_{i,j} \frac{\gamma_{ij} }{2} 
\pc{2 \sigma^-_i \rho \sigma^+_j - \sigma^+_i \sigma_j^- \rho - \rho \sigma^+_i \sigma^-_j }.
\end{align}
The parameters $\gamma_{ij}$ and $g_{ij}$ can be written compactly as
\begin{align}
g_{ij} &=  \frac{3 \Gamma}{2} 
\left\lbrace
- \pc{ 1-\cos^2 \theta_{dd} } \frac{\cos \pc{k_0 r_{ij}} }{k_0 r_{ij}}
\right. \nonumber \\
&\quad + 
\left. \pc{ 1-3 \cos^2 \theta_{dd}} \pr{ \frac{\sin \pc{k_0 r_{ij}}}{\pc{k_0 r_{ij}}^2} + \frac{\cos \pc{k_0 r_{ij}}}{\pc{k_0 r_{ij}}^3} }
\right\rbrace, 
\label{eq:apg12}
\end{align}
\begin{align}
\gamma_{ij} &=  \frac{3 \Gamma}{2}  
\left\lbrace
\pc{ 1-\cos^2 \theta_{dd} } \frac{\sin \pc{k_0 r_{ij}}}{k_0 r_{ij}} \right. \nonumber \\
&\quad +
\left.
\pc{ 1-3 \cos^2 \theta_{dd}} \pr{ \frac{\cos \pc{k_0 r_{ij}}}{\pc{k_0 r_{ij}}^2}-\frac{\sin \pc{k_0 r_{ij}}}{\pc{k_0 r_{ij}}^3} }
\right\rbrace, \label{eq:apgamma12}
\end{align}
where $\cos \theta_{dd} = e_{\mathbf{d}} \cdot e_{\mathbf{r}_{ij}}$. 

\section{Solving the dynamics of a two-atom system driven by a coherent laser field \label{append:dynamics}}

Consider two identical atoms with levels $\ket{e}_j$ and $\ket{g}_j$ $(j=1,2)$ at fixed positions $\mathbf{r}_1$ and $\mathbf{r}_2$ (separated by a distance $r_{12}$) with dipole moment $\mathbf{d}_{eg}$ and transition frequency $\omega_0$, as described in Sec.~\ref{sec:Theory}. They are driven by a resonant external laser field with wavevector $\mathbf{k}_L$. The time evolution of the reduced atomic density operator $\rho$ in the laboratory frame is given by the master equation \cite{PRA52.636(1995), PhysRep372.369(2002)}
\begin{align}
\frac{\partial \rho }{\partial t} &= -\frac{\mathrm{i}}{\hbar }\pr{\hH,\, \rho} 
\nonumber \\
&\quad
-\sum _{i,j=1}^2 \gamma _{ij} \left(\sigma_i^+ \sigma_j^- \, \rho  + \rho \, \sigma_i^+ \sigma_j^- -2  \, \sigma_j^- \, \rho \, \sigma_i^+\right),
\end{align}
where we assume that the only dissipative terms are due to the spontaneous decays of the levels $\ket{e}_{1,2}$ and $\gamma_{ij} \,(i\neq j)$ is given by Eq.~\eqref{eq:apgamma12} and $\gamma_{ii} = \Gamma$. 
The Hamiltonian $\hH$ of the system is given by
\begin{align}
\hH = \hH_0 +\hH_\text{int}+\hH_{L},
\end{align}
where
\begin{align}
\hH_0 = \hbar \omega_0 \sum_{i=1}^2 \sigma_i^z
\end{align}
is the unperturbed Hamiltonian of the atoms with $\sigma_i^z$ describing their energy. The second term represents the dipole-dipole interaction between the atoms
\begin{align}
\hH_\text{int} =\frac{1}{2} \hbar \sum_{i \neq j} g_{ij} \pc{\sigma_i^+ \sigma_j^- + \text{H.c.}}
\end{align}
where $g_{ij}$ is defined by eq.~\eqref{eq:apg12}.
Finally, the third term describes the coupling between the driving field and the atoms, which in the rotating-wave approximation is given by
\begin{align}
\hH_L =-\frac{\hbar}{2} \sum_{i=1}^2 \pr{\Omega_i \, \sigma_i^+ \exp \pc{\mathrm{i} \omega_L t} + \text{H.c.}},
\end{align}
where $\omega_L$ is the laser frequency, and $\Omega_i$ is the Rabi frequency of the driving field at the position of the $i$th atom.

To study the dynamics of the system, we need to determine, from the master equation given, correlation functions for the atomic operators, for any atomic operator $Q$, in terms of the reduced density operator $\rho$ \cite{PRA52.636(1995)}
\begin{align}
\av{Q} = \text{Tr}_S \pr{\rho \, Q}
\end{align}
where the trace is only over the atoms plus the laser mode subsystem. For the two-atom system, substituting the atomic operators $\sigma_i^\pm \; (i=1,2)$ for $O$ and using the master equation, it is possible to obtain a closed system of 15 first-order differential equations,
\begin{align*}
S_1(t) &= \av{\tilde{\sigma}^+_1 (t)},\\
S_2 (t) &=\av{\tilde{\sigma}^-_1 (t)},\\
S_3 (t) &= \av{\tilde{\sigma}^+_2 (t)},\\
S_4 (t) &= \av{\tilde{\sigma}^-_2 (t)},\\
S_5 (t) &= \av{\tilde{\sigma}^+_1 (t) \, \tilde{\sigma}^-_1 (t)},\\
S_6 (t) &=\av{\tilde{\sigma}^+_2 (t)\, \tilde{\sigma}^-_2 (t)},\\
S_7 (t) &= \av{\tilde{\sigma}^+_1 (t) \, \tilde{\sigma}^-_2 (t)},\\
S_8 (t) &= \av{\tilde{\sigma}^+_2 (t)\, \tilde{\sigma}^-_1 (t)},\\
S_9 (t) &= \av{\tilde{\sigma}^+_1 (t) \, \tilde{\sigma}^+_2 (t)},\\
S_{10} (t) &= \av{\tilde{\sigma}^-_1 (t)\, \tilde{\sigma}^-_2 (t)},\\
S_{11} (t) &=\av{\tilde{\sigma}^+_1 (t) \, \tilde{\sigma}^-_1 (t) \tilde{\sigma}_2^- (t) }, \\
S_{12} (t) &=\av{\tilde{\sigma}^+_1 (t) \, \tilde{\sigma}^+_2 (t) \,  \tilde{\sigma}_1^- (t) }, \\
S_{13} (t) &=\av{\tilde{\sigma}^+_2 (t) \, \tilde{\sigma}^-_1 (t) \, \tilde{\sigma}_2^- (t) }, \\
S_{14} (t) &= \av{\tilde{\sigma}^+_1 (t) \, \tilde{\sigma}^+_2 (t)\, \tilde{\sigma}_2^- (t) }, \\
S_{15} (t) &=\av{\tilde{\sigma}^+_1 (t) \, \tilde{\sigma}^+_2 (t) \, \tilde{\sigma}^-_1 (t) \, \tilde{\sigma}_2^- (t) },
\end{align*} 
where
\begin{align}
\tilde{\sigma}^\pm_i (t) = \sigma^\pm_i \exp \pc{\mp \mathrm{i} \omega_L t}.
\end{align}
We assume that $\pc{\sigma^\pm_i}^2 = 0$ and that the atomic operators for different atoms commute at the same time,
\begin{align}
\sigma_i^\pm \sigma^\pm_j = \sigma_j^\pm \sigma_i^\pm
\quad \quad \text{and} \quad \quad \sigma_i^\pm \sigma_j^\mp = \sigma_j^\mp \sigma_i^\pm \quad \text{for} \; i \neq j.
\end{align}
Finally, the 15 equations of motion become
\begin{widetext}
\begin{align}
\dot{S}_1 &= -\frac{\mathrm{i} \Omega_1^* }{2}+2 S_{12} (\gamma_{12}-\mathrm{i} g_{12})-S_3 (\gamma_{12}-\mathrm{i} g_{12})+(-(1+\mathrm{i} \Delta_1 )) S_1 + \mathrm{i} \Omega_1^* S_5,\\
\dot{S}_2 &= \frac{\mathrm{i} \Omega_1 }{2}+2 S_{11} (\gamma_{12}+\mathrm{i} g_{12})- S_4 (\gamma_{12}+\mathrm{i} g_{12})+(-(1-\mathrm{i} \Delta_1 )) S_2-\mathrm{i} \Omega_1 S_5,\\
\dot{S}_3 &=-\frac{\mathrm{i} \Omega_2^*}{2}+\mathrm{i} S_6 \Omega_2^*+ S_1 (-(\gamma_{12}-\mathrm{i} g_{12}))+2 S_{14} (\gamma_{12}-\mathrm{i} g)+(-(1+\mathrm{i} \Delta_2 )) S_3,\\
\dot{S}_4 &=\frac{\mathrm{i} \Omega_2}{2}+2 S_{13} (\gamma_{12}+\mathrm{i} g_{12})- S_2 (\gamma_{12}+\mathrm{i} g_{12})+(-(1-\mathrm{i} \Delta_2 )) S_4-\mathrm{i} \Omega_2 S_6,\\
\dot{S}_5 &= -S_7 (\gamma_{12}+\mathrm{i} g_{12}) - S_8 (\gamma_{12}-\mathrm{i} g_{12})+\frac{\mathrm{i} \Omega_1  S_1}{2}-\frac{\mathrm{i} \Omega_1^* S_2}{2}-2 S_5,\\
\dot{S}_6 &=-\frac{\mathrm{i} S_4 \Omega_2^*}{2}-S_7 (\gamma_{12}-\mathrm{i} g_{12})-S_8 (\gamma_{12}+\mathrm{i} g_{12})+\frac{\mathrm{i} \Omega_2 S_3}{2}-2 S_6,\\
\dot{S}_7 &= S_5 (-(\gamma_{12}+\mathrm{i} g_{12}))- S_6 (\gamma_{12}-\mathrm{i} g_{12})+\frac{\mathrm{i}\Omega_2 S_1}{2}+\mathrm{i} \Omega_1^*  S_{11}-\mathrm{i} \Omega_2 S_{14}  + 2 \gamma_{12} S_{15}-\frac{\mathrm{i} \Omega_1^* S_4}{2}-(2 +\mathrm{i} \Delta_1 -\mathrm{i}\Delta_2) S_7,\\
\dot{S}_8 &= \mathrm{i} S_{13} \Omega_2^*-\frac{\mathrm{i} S_2 \Omega_2^*}{2}+S_5 (-(\gamma_{12}-\mathrm{i} g_{12}))-S_6 (\gamma_{12}+\mathrm{i} g_{12})-\mathrm{i} \Omega_1 S_{12}
 +2 \gamma_{12} S_{15}+\frac{\mathrm{i} \Omega_1  S_3}{2}-(2 - \mathrm{i}\Delta_1 + \mathrm{i} \Delta_2) S_8,\\
\dot{S}_9 &=-\frac{\mathrm{i} S_1 \Omega_2^*}{2}+\mathrm{i} S_{14} \Omega_2^*+\mathrm{i} \Omega_1^* S_{12}-\frac{\mathrm{i} \Omega_1^* S_3}{2}-2 \pc{1+\mathrm{i} \frac{\Delta_1 +\Delta_2}{2} } S_9,\\
\dot{S}_{10} &=-2 \pc{1-\mathrm{i} \frac{\Delta_1 +\Delta_2}{2} } S_{10}-\mathrm{i} \Omega_1 S_{11}-\mathrm{i} \Omega_2 S_{13}+\frac{\mathrm{i} \Omega_2 S_2}{2}+\frac{\mathrm{i} \Omega_1 S_4}{2},\\
\dot{S}_{11} &= S_{13} (-(\gamma_{12}-\mathrm{i} g_{12}))-\frac{\mathrm{i} \Omega_1^* S_{10}}{2}+(-(3-\mathrm{i} \Delta_2 )) S_{11}-\mathrm{i} \Omega_2 S_{15}+\frac{\mathrm{i} \Omega_2 S_5}{2}+\frac{\mathrm{i} \Omega_1 S_7}{2},\\
\dot{S}_{12} &=\mathrm{i} S_{15} \Omega_2^*-\frac{\mathrm{i} S_5 \Omega_2^*}{2}+S_{14} (-(\gamma_{12}+\mathrm{i} g_{12}))+(-(3+\mathrm{i} \Delta_2 )) S_{12}-\frac{\mathrm{i} \Omega_1^* S_8}{2}+\frac{\mathrm{i} \Omega_1  S_9}{2}, \\
\dot{S}_{13} &=-\frac{\mathrm{i} S_{10} \Omega_2^*}{2} + S_{11} (-(\gamma_{12}-\mathrm{i} g_{12}))+(-(3-\mathrm{i} \Delta_1 )) S_{13}-\mathrm{i} \Omega_1 S_{15}+\frac{\mathrm{i} \Omega_1 S_6}{2}+\frac{\mathrm{i} \Omega_2 S_8}{2}\\
\dot{S}_{14} &= -\frac{\mathrm{i} S_7 \Omega_2^*}{2} + S_{12} (-(\gamma_{12}+\mathrm{i} g_{12}))+(-(3+\mathrm{i} \Delta_1 )) S_{14}+\mathrm{i} \Omega_1^* S_{15}-\frac{\mathrm{i} \Omega_1^* S_6}{2}+\frac{\mathrm{i} \Omega_2 S_9}{2},\\
\dot{S}_{15} &= -\frac{\mathrm{i} S_{11} \Omega_2^*}{2}+\frac{\mathrm{i} \Omega_2 S_{12}}{2}-\frac{\mathrm{i} \Omega_1^*  S_{13}}{2}+\frac{\mathrm{i} \Omega_1 S_{14}}{2}-4 S_{15}.
\end{align}

It is now possible to solve the set of equations for the steady state and study the photon statistics of our systems.

\section{Solving second-order multi-time correlation function $G^{(2)} (\mathbf{R};0, t)$ \label{App:G2}}

As is referred to in the main text, the solution to the master equation only yields single time averages $$\dot{\mathbf{S} }(t) = \mathbf{M}\mathbf{S} (t)  + \mathbf{b}.$$ To find the equation of motion for multi-time vectors, we need to take the single-time equation of motion, multiplying on the left-hand side by $\sigma^+_i (0)$ and on the right by $\sigma^-_j (0)$ \cite{BookCarmichael}:  
    \begin{align*}
        \frac{d }{dt} \av{\sigma^+_i (0) \, \mathbf{S} (t) \,  \sigma^-_j (0)} &= \mathbf{M}\av{\sigma^+_i (0) \, \mathbf{S} (t) \, \sigma^-_j (0)}  + \av{\sigma^+_i (0)\, \sigma^-_j (0)}\mathbf{b},\\
        &= \mathbf{M} \pr{ \av{\sigma^+_i (0) \, \mathbf{S} (t) \, \sigma^-_j (0)}  + \av{\sigma^+_i (0) \, \sigma^-_j (0)} \mathbf{M}^{-1} \mathbf{b}}.
    \end{align*}
The formal solution to this equation is then given by 
    \begin{align*}
        \av{\sigma^+_i (0) \, \mathbf{S} (t) \, \sigma^-_j (0)} &=
        - \av{\sigma^+_i \sigma^-_j}_\text{ss} \mathbf{M}^{-1} \mathbf{b} + \exp (\mathbf{M} t) 
        \pr{\av{\sigma^+_i \, \mathbf{S} \, \sigma^-_j }_\text{ss}   + \av{\sigma^+_i \sigma^-_j}_\text{ss} \mathbf{M}^{-1} \mathbf{b} } \\
        &= \exp (\mathbf{M} t) \av{\sigma^+_i \, \mathbf{S} \, \sigma^-_j }_\text{ss} + \av{\sigma^+_i \sigma^-_j}_\text{ss} 
        \pr{ \exp (\mathbf{M} t) - 1} \mathbf{M}^{-1} \mathbf{b} .
    \end{align*}
\end{widetext}
To find the vectors corresponding to $\av{\sigma^+_i \, \mathbf{S} \, \sigma^-_j}_\text{ss}$ we assume once again that $\pc{\sigma^\pm_i}^2 = 0$ and that the atomic operators for different atoms commute at the same time. For $i=1, \, j=2$, we find
\begin{align*}
\av{ \sigma^+_1 S_1 \sigma_2^-}_\text{ss} &= \av{\sigma^+_1 \tilde{\sigma}^+_1 \sigma_2^-}_\text{ss} =0 ,\\
\av{ \sigma^+_1 S_2  \sigma_2^-}_\text{ss} &=\av{\sigma^+_1 \tilde{\sigma}^-_1 \sigma_2^-}_\text{ss} = S_{11}^\text{ss} ,\\
\av{ \sigma^+_1 S_3 \sigma_2^-}_\text{ss} &= \av{\sigma^+_1 \tilde{\sigma}^+_2 \sigma_2^-}_\text{ss} =  S_{14}^\text{ss},\\
\av{ \sigma^+_1 S_4  \sigma_2^-}_\text{ss} &= \av{\sigma^+_1 \tilde{\sigma}^-_2 \sigma_2^-}_\text{ss} =0 ,\\
\av{ \sigma^+_1 S_5 \sigma_2^-}_\text{ss} &= \av{\sigma^+_1 \tilde{\sigma}^+_1  \tilde{\sigma}^-_1 \sigma_2^-}_\text{ss} =0,\\
\av{ \sigma^+_1 S_6 \sigma_2^-}_\text{ss}&=\av{\sigma^+_1 \tilde{\sigma}^+_2 \tilde{\sigma}^-_2 \sigma_2^-}_\text{ss} =0 ,\\
\av{ \sigma^+_1 S_7 \sigma_2^-}_\text{ss} &= \av{\sigma^+_1 \tilde{\sigma}^+_1 \tilde{\sigma}^-_2 \sigma_2^-}_\text{ss}=0,\\
\av{ \sigma^+_1 S_8 \sigma_2^-}_\text{ss} &= \av{\sigma^+_1 \tilde{\sigma}^+_2 \tilde{\sigma}^-_1 \sigma_2^-}_\text{ss} = S_{15}^\text{ss} ,\\
\av{ \sigma^+_1 S_9 \sigma_2^-}_\text{ss} &= \av{\sigma^+_1 \tilde{\sigma}^+_1  \tilde{\sigma}^+_2\sigma_2^-}_\text{ss} =0 ,\\ 
\av{ \sigma^+_1 S_{10} \sigma_2^-}_\text{ss} &= \av{\sigma^+_1 \tilde{\sigma}^-_1 \tilde{\sigma}^-_2 \sigma_2^-}_\text{ss} =0,\\
\av{ \sigma^+_1 S_{11} \sigma_2^-}_\text{ss} &=\av{\sigma^+_1 \tilde{\sigma}^+_1 \tilde{\sigma}^-_1  \tilde{\sigma}_2^-  \sigma_2^-}_\text{ss}=0, \\
\av{ \sigma^+_1 S_{12} \sigma_2^-}_\text{ss} &=\av{\sigma^+_1 \tilde{\sigma}^+_1  \tilde{\sigma}^+_2  \tilde{\sigma}_1^- \sigma_2^-}_\text{ss}=0, \\
\av{ \sigma^+_1  S_{13} \sigma_2^-}_\text{ss}  &=\av{\sigma^+_1 \tilde{\sigma}^+_2 \tilde{\sigma}^-_1  \tilde{\sigma}_2^-  \sigma_2^-}_\text{ss} =0 , \\
\av{ \sigma^+_1 S_{14} \sigma_2^-}_\text{ss} &= \av{\sigma^+_1 \tilde{\sigma}^+_1  \tilde{\sigma}^+_2  \tilde{\sigma}_2^-  \sigma_2^-}_\text{ss}=0, \\
\av{ \sigma^+_1  S_{15} \sigma_2^-}_\text{ss} &=\av{\sigma^+_1 \tilde{\sigma}^+_1  \tilde{\sigma}^+_2  \tilde{\sigma}^-_1 \tilde{\sigma}_2^- \sigma_2^-}_\text{ss} =0.
\end{align*} 

For $i=2, \, j=1$,
\begin{align*}
\av{ \sigma^+_2 S_1 \sigma_1^-}_\text{ss} &= \av{\sigma^+_2 \tilde{\sigma}^+_1 \sigma_1^-}_\text{ss} = S_{12}^\text{ss}  ,\\
\av{ \sigma^+_2 S_2  \sigma_1^-}_\text{ss} &=\av{\sigma^+_2 \tilde{\sigma}^-_1 \sigma_1^-}_\text{ss} = 0,\\
\av{ \sigma^+_2 S_3 \sigma_1^-}_\text{ss} &= \av{\sigma^+_2 \tilde{\sigma}^+_2 \sigma_1^-}_\text{ss} =  0,\\
\av{ \sigma^+_2 S_4  \sigma_1^-}_\text{ss} &= \av{\sigma^+_2 \tilde{\sigma}^-_2 \sigma_1^-}_\text{ss} = S_{13}^\text{ss} ,\\
\av{ \sigma^+_2 S_5 \sigma_1^-}_\text{ss} &= \av{\sigma^+_2 \tilde{\sigma}^+_1  \tilde{\sigma}^-_1 \sigma_1^-}_\text{ss} =0,\\
\av{ \sigma^+_2 S_6 \sigma_1^-}_\text{ss}&=\av{\sigma^+_2 \tilde{\sigma}^+_2 \tilde{\sigma}^-_2 \sigma_1^-}_\text{ss} =0 ,\\
\av{ \sigma^+_2 S_7 \sigma_1^-}_\text{ss} &= \av{\sigma^+_2 \tilde{\sigma}^+_1 \tilde{\sigma}^-_2 \sigma_1^-}_\text{ss}=S_{15}^\text{ss},\\
\av{ \sigma^+_2 S_8 \sigma_1^-}_\text{ss} &= \av{\sigma^+_2 \tilde{\sigma}^+_2 \tilde{\sigma}^-_1 \sigma_1^-}_\text{ss} =0,\\
\av{ \sigma^+_2 S_9 \sigma_1^-}_\text{ss} &= \av{\sigma^+_2 \tilde{\sigma}^+_1  \tilde{\sigma}^+_2\sigma_1^-}_\text{ss} =0 ,\\ 
\av{ \sigma^+_2 S_{10} \sigma_1^-}_\text{ss} &= \av{\sigma^+_2 \tilde{\sigma}^-_1 \tilde{\sigma}^-_2 \sigma_1^-}_\text{ss} =0,\\
\av{ \sigma^+_2 S_{11} \sigma_1^-}_\text{ss} &=\av{\sigma^+_2 \tilde{\sigma}^+_1 \tilde{\sigma}^-_1  \tilde{\sigma}_2^-  \sigma_1^-}_\text{ss}=0, \\
\av{ \sigma^+_2 S_{12} \sigma_1^-}_\text{ss} &=\av{\sigma^+_2 \tilde{\sigma}^+_1  \tilde{\sigma}^+_2  \tilde{\sigma}_1^- \sigma_1^-}_\text{ss}=0, \\
\av{ \sigma^+_2  S_{13} \sigma_1^-}_\text{ss}  &=\av{\sigma^+_2 \tilde{\sigma}^+_2 \tilde{\sigma}^-_1  \tilde{\sigma}_2^-  \sigma_1^-}_\text{ss} =0 , \\
\av{ \sigma^+_2 S_{14} \sigma_1^-}_\text{ss} &= \av{\sigma^+_2 \tilde{\sigma}^+_1  \tilde{\sigma}^+_2  \tilde{\sigma}_2^-  \sigma_1^-}_\text{ss}=0, \\
\av{ \sigma^+_2  S_{15} \sigma_1^-}_\text{ss} &=\av{\sigma^+_2 \tilde{\sigma}^+_1  \tilde{\sigma}^+_2  \tilde{\sigma}^-_1 \tilde{\sigma}_2^- \sigma_1^-}_\text{ss} =0.
\end{align*} 

For $i, j=1$,
\begin{align*}
\av{ \sigma^+_1 S_1 \sigma_1^-}_\text{ss} &= \av{\sigma^+_1 \tilde{\sigma}^+_1 \sigma_1^-}_\text{ss} =0 ,\\
\av{ \sigma^+_1 S_2  \sigma_1^-}_\text{ss} &=\av{\sigma^+_1 \tilde{\sigma}^-_1 \sigma_1^-}_\text{ss} = 0,\\
\av{ \sigma^+_1 S_3 \sigma_1^-}_\text{ss} &= \av{\sigma^+_1 \tilde{\sigma}^+_2 \sigma_1^-}_\text{ss} =  S_{12}^\text{ss},\\
\av{ \sigma^+_1 S_4  \sigma_1^-}_\text{ss} &= \av{\sigma^+_1 \tilde{\sigma}^-_2 \sigma_1^-}_\text{ss} = S_{11}^\text{ss},\\
\av{ \sigma^+_1 S_5 \sigma_1^-}_\text{ss} &= \av{\sigma^+_1 \tilde{\sigma}^+_1  \tilde{\sigma}^-_1 \sigma_1^-}_\text{ss} =0,\\
\av{ \sigma^+_1 S_6 \sigma_1^-}_\text{ss}&=\av{\sigma^+_1 \tilde{\sigma}^+_2 \tilde{\sigma}^-_2 \sigma_1^-}_\text{ss} = S_{15}^\text{ss} ,\\
\av{ \sigma^+_1 S_7 \sigma_1^-}_\text{ss} &= \av{\sigma^+_1 \tilde{\sigma}^+_1 \tilde{\sigma}^-_2 \sigma_1^-}_\text{ss}=0,\\
\av{ \sigma^+_1 S_8 \sigma_1^-}_\text{ss} &= \av{\sigma^+_1 \tilde{\sigma}^+_2 \tilde{\sigma}^-_1 \sigma_1^-}_\text{ss} =0,\\
\av{ \sigma^+_1 S_9 \sigma_1^-}_\text{ss} &= \av{\sigma^+_1 \tilde{\sigma}^+_1  \tilde{\sigma}^+_2\sigma_1^-}_\text{ss} =0 ,\\ 
\av{ \sigma^+_1 S_{10} \sigma_1^-}_\text{ss} &= \av{\sigma^+_1 \tilde{\sigma}^-_1 \tilde{\sigma}^-_2 \sigma_1^-}_\text{ss} =0,\\
\av{ \sigma^+_1 S_{11} \sigma_1^-}_\text{ss} &=\av{\sigma^+_1 \tilde{\sigma}^+_1 \tilde{\sigma}^-_1  \tilde{\sigma}_2^-  \sigma_1^-}_\text{ss}=0, \\
\av{ \sigma^+_1 S_{12} \sigma_1^-}_\text{ss} &=\av{\sigma^+_1 \tilde{\sigma}^+_1  \tilde{\sigma}^+_2  \tilde{\sigma}_1^- \sigma_1^-}_\text{ss}=0, \\
\av{ \sigma^+_1  S_{13} \sigma_1^-}_\text{ss}  &=\av{\sigma^+_1 \tilde{\sigma}^+_2 \tilde{\sigma}^-_1  \tilde{\sigma}_2^-  \sigma_1^-}_\text{ss} =0 , \\
\av{ \sigma^+_1 S_{14} \sigma_1^-}_\text{ss} &= \av{\sigma^+_1 \tilde{\sigma}^+_1  \tilde{\sigma}^+_2  \tilde{\sigma}_2^-  \sigma_1^-}_\text{ss}=0, \\
\av{ \sigma^+_1  S_{15} \sigma_1^-}_\text{ss} &=\av{\sigma^+_1 \tilde{\sigma}^+_1  \tilde{\sigma}^+_2  \tilde{\sigma}^-_1 \tilde{\sigma}_2^- \sigma_1^-}_\text{ss} =0.
\end{align*} 

For $i,j=2$,
\begin{align*}
\av{ \sigma^+_2 S_1 \sigma_2^-}_\text{ss} &= \av{\sigma^+_2 \tilde{\sigma}^+_1 \sigma_2^-}_\text{ss} = S_{14}^\text{ss},\\
\av{ \sigma^+_2 S_2  \sigma_2^-}_\text{ss} &=\av{\sigma^+_2 \tilde{\sigma}^-_1 \sigma_2^-}_\text{ss} = S_{13}^\text{ss},\\
\av{ \sigma^+_2 S_3 \sigma_2^-}_\text{ss} &= \av{\sigma^+_2 \tilde{\sigma}^+_2 \sigma_2^-}_\text{ss} =  0,\\
\av{ \sigma^+_2 S_4  \sigma_2^-}_\text{ss} &= \av{\sigma^+_2 \tilde{\sigma}^-_2 \sigma_2^-}_\text{ss} =0 ,\\
\av{ \sigma^+_2 S_5 \sigma_2^-}_\text{ss} &= \av{\sigma^+_2 \tilde{\sigma}^+_1  \tilde{\sigma}^-_1 \sigma_2^-}_\text{ss} = S_{15}^\text{ss},\\
\av{ \sigma^+_2 S_6 \sigma_2^-}_\text{ss}&=\av{\sigma^+_2 \tilde{\sigma}^+_2 \tilde{\sigma}^-_2 \sigma_2^-}_\text{ss} =0 ,\\
\av{ \sigma^+_2 S_7 \sigma_2^-}_\text{ss} &= \av{\sigma^+_2 \tilde{\sigma}^+_1 \tilde{\sigma}^-_2 \sigma_2^-}_\text{ss}=0,\\
\av{ \sigma^+_2 S_8 \sigma_2^-}_\text{ss} &= \av{\sigma^+_2 \tilde{\sigma}^+_2 \tilde{\sigma}^-_1 \sigma_2^-}_\text{ss} =0,\\
\av{ \sigma^+_2 S_9 \sigma_2^-}_\text{ss} &= \av{\sigma^+_2 \tilde{\sigma}^+_1  \tilde{\sigma}^+_2\sigma_2^-}_\text{ss} =0 ,\\ 
\av{ \sigma^+_2 S_{10} \sigma_2^-}_\text{ss} &= \av{\sigma^+_2 \tilde{\sigma}^-_1 \tilde{\sigma}^-_2 \sigma_2^-}_\text{ss} =0, \\
\av{ \sigma^+_2 S_{11} \sigma_2^-}_\text{ss} &=\av{\sigma^+_2 \tilde{\sigma}^+_1 \tilde{\sigma}^-_1  \tilde{\sigma}_2^-  \sigma_2^-}_\text{ss}=0, \\
\av{ \sigma^+_2 S_{12} \sigma_2^-}_\text{ss} &=\av{\sigma^+_2 \tilde{\sigma}^+_1  \tilde{\sigma}^+_2  \tilde{\sigma}_1^- \sigma_2^-}_\text{ss}=0, \\
\av{ \sigma^+_2  S_{13} \sigma_2^-}_\text{ss}  &=\av{\sigma^+_2 \tilde{\sigma}^+_2 \tilde{\sigma}^-_1  \tilde{\sigma}_2^-  \sigma_2^-}_\text{ss} =0 , \\
\av{ \sigma^+_2 S_{14} \sigma_2^-}_\text{ss} &= \av{\sigma^+_2 \tilde{\sigma}^+_1  \tilde{\sigma}^+_2  \tilde{\sigma}_2^-  \sigma_2^-}_\text{ss}=0, \\
\av{ \sigma^+_2  S_{15} \sigma_2^-}_\text{ss} &=\av{\sigma^+_2 \tilde{\sigma}^+_1  \tilde{\sigma}^+_2  \tilde{\sigma}^-_1 \tilde{\sigma}_2^- \sigma_2^-}_\text{ss} =0.
\end{align*} 
This calculation is independent of the form of the matrix $\mathbf{M}$ and $\mathbf{b}$ and we can calculate $G^{(2)} (\mathbf{R};0, t)$ for the different experimental parameters with this recipe.




\begin{thebibliography}{99}

\bibitem{Nature177.27(1956)}
R. Hanbury Brown and R. Q. Twiss,
\href{https://doi.org/10.1038/177027a0}{Nature \textbf{177}, 27-29 (1956).}

\bibitem{Nature178.1046(1956)}
R. Hanbury Brown and R. Q. Twiss,
\href{https://doi.org/10.1038/1781046a0}{Nature \textbf{178}, 1046-1048 (1956)}.

\bibitem{NatPhot3.696(2009)}
Robert H. Hadfield,
\href{https://doi.org/10.1038/nphoton.2009.230}{Nat. Photonics \textbf{3}, 696-705 (2009)}.

\bibitem{RevScienInst82.071101(2011)}
M. D. Eisaman, J. Fan, A. Migdall, and S. V. Polyakov,
\href{https://doi.org/10.1063/1.3610677}{Rev. Sci. Instrum. \textbf{82}, 071101 (2011)}.

\bibitem{OptPhotNews30.32(2019)}
Urbasi Sinha, Surya Narayan Sahoo, Ashutosh Singh, Kaushik Joarder, Rishab Chatterjee, and Sanchari Chakraborti,
\href{https://doi.org/10.1364/OPN.30.9.000032}{Opt. Photonics News \textbf{30}(9), 32-39 (2019)}.

\bibitem{PRL74.3600(1995)}
D. V. Strekalov, A. V. Sergienko, D. N. Klyshko, and Y. H. Shih,
\href{https://doi.org/10.1103/PhysRevLett.74.3600}{Phys. Rev. Lett. \textbf{74}, 3600 (1995)}.

\bibitem{PRA70.051802(R)(2004)}
Giuliano Scarcelli, Alejandra Valencia, and Yanhua Shih,
\href{https://doi.org/10.1103/PhysRevA.70.051802}{Phys. Rev. A \textbf{70}, 051802(R) (2004)}.

\bibitem{PRA73.053802(2006)}
M. Bache, D. Magatti, F. Ferri, A. Gatti, E. Brambilla, and L. A. Lugiato,
\href{https://doi.org/10.1103/PhysRevA.73.053802}{Phys. Rev. A \textbf{73}, 053802 (2006)}.

\bibitem{PRL119.263603(2017)}
Yong Sup Ihn, Yosep Kim, Vincenzo Tamma, and Yoon-Ho Kim,
\href{https://doi.org/10.1103/PhysRevLett.119.263603}{Phys. Rev. Lett. \textbf{119}, 263603 (2017)}.

\bibitem{PRL16.1012(1966)}
B. L. Morgan and L. Mandel,
\href{https://doi.org/10.1103/PhysRevLett.16.1012}{Phys. Rev. Lett. \textbf{16}, 1012 (1966)}.

\bibitem{PRA53.3469(1996)}
S. Bali, D. Hoffmann, J. Sim\'an, and T. Walker,
\href{https://doi.org/10.1103/PhysRevA.53.3469}{Phys. Rev. A \textbf{53}, 3469 (1996)}.

\bibitem{OptLett29.2713(2004)}
R. Stites, M. Beeler, L. Feeney, S. Kim, and S. Bali,
\href{https://doi.org/10.1364/OL.29.002713}{Opt. Lett. \textbf{29}, 2713-2715 (2004)}.

\bibitem{OptExp18.6604(2010)}
K. Nakayama, Y. Yoshikawa, H. Matsumoto, Y. Torii, and T. Kuga,
\href{https://doi.org/10.1364/OE.18.006604}{Opt. Exp. \textbf{18}, 6604-6612 (2010)}.

\bibitem{PRA93.043826(2016)}
A. Dussaux, T. Passerat de Silans, W. Guerin, O. Alibart, S. Tanzilli, F. Vakili, and R. Kaiser,
\href{https://doi.org/10.1103/PhysRevA.93.043826}{Phys. Rev. A \textbf{93}, 043826 (2016)}.

\bibitem{PRB78.153309(2008)}
Ahsan Nazir,
\href{https://doi.org/10.1103/PhysRevB.78.153309}{Phys. Rev. B \textbf{78}, 153309 (2008)}.

\bibitem{JOSA61.1307(1971)}
M. Rousseau,
\href{https://doi.org/10.1364/JOSA.61.001307}{J. Opt. Soc. Am. \textbf{61}, 1307-1316 (1971)}.

\bibitem{PRL116.050401(2016)}
Mihai D. Vidrighin, Oscar Dahlsten, Marco Barbieri, M. S. Kim, Vlatko Vedral, and Ian A. Walmsley,
\href{https://doi.org/10.1103/PhysRevLett.116.050401}{Phys. Rev. Lett. \textbf{116}, 050401 (2016)}.

\bibitem{JPhysB49.025301(2016)}
K. Muhammed Shafi, Deepak Pandey , Buti Suryabrahmam, B. S. Girish, and Hema Ramachandran,
\href{https://iopscience.iop.org/article/10.1088/0953-4075/49/2/025301}{J. Phys. B: At. Mol. Opt. Phys. \textbf{49}, 025301 (2016)}.

\bibitem{NJP20.093002(2018)}
J. Mika, L. Podhora , L. Lachman, P. Ob\v{s}il, J. Hlou\v{s}ek, M. Je\v{z}ek, R. Filip, and L. Slodi\v{c}ka,
\href{https://doi.org/10.1088/1367-2630/aadc9d}{New J. Phys. \textbf{20}, 093002 (2018)}.

\bibitem{ScienRep8.10981(2018)}
Jiho Park, Taek Jeong and Han Seb Moon,
\href{https://doi.org/10.1038/s41598-018-29340-7}{Sci. Rep. \textbf{8}, 10981 (2018)}.

\bibitem{PRL92.213601(2004)}
C. W. Chou, S. V. Polyakov, A. Kuzmich, and H. J. Kimble,
\href{https://doi.org/10.1103/PhysRevLett.92.213601}{Phys. Rev. Lett. \textbf{92}, 213601 (2004)}.

\bibitem{Science336.887(2012)}
Y. O. Dudin, A. Kuzmich,
\href{https://doi.org/10.1126/science.1217901}{Science \textbf{336}, 887-889 (2012)}.

\bibitem{NJP11.103004(2009)}
H. G. Barros, A. Stute, T. E. Northup, C. Russo, P. O. Schmidt and R. Blatt,
\href{https://doi.org/10.1088/1367-2630/11/10/103004}{New J. Phys. \textbf{11}, 103004 (2009)}.

\bibitem{PRL83.2722(1999)}
C. Brunel, B. Lounis, P. Tamarat and M. Orrit,
\href{https://doi.org/10.1103/PhysRevLett.83.2722}{Phys. Rev. Lett. \textbf{83}, 2722-2725 (1999)}.

\bibitem{Science298.385(2002)}
C. Hettich, C. Schmitt, J. Zitzmann, S. K\"uhn, I. Gerhardt, and V. Sandoghdar,
\href{https://science.sciencemag.org/content/298/5592/385}{Science \textbf{298}, 385-389 (2002).}

\bibitem{npjQuantInf(2018)}
Lukas Hanschke, Kevin A. Fischer, Stefan Appel, Daniil Lukin, Jakob Wierzbowski, Shuo Sun, Rahul Trivedi, Jelena Vu\v{c}kovi\'c, Jonathan J. Finley and Kai M\"uller,
\href{https://doi.org/10.1038/s41534-018-0092-0}{npj Quantum Inf. \textbf{4}, 43 (2018)}.

\bibitem{PRL85.290(2000)}
Christian Kurtsiefer, Sonja Mayer, Patrick Zarda, and Harald Weinfurter,
\href{https://doi.org/10.1103/PhysRevLett.85.290}{Phys. Rev. Lett. \textbf{85}, 290 (2000)}.

\bibitem{PRA87.053412(2013)}
M. M. M\"uller, A. K\"olle, R. L\"ow, T. Pfau, T. Calarco, and S. Montangero,
\href{https://doi.org/10.1103/PhysRevA.87.053412}{Phys. Rev. A \textbf{87}, 053412 (2013)}.

\bibitem{PRL118.253602(2017)}
D. J. Whiting, N. Sibali\'c, J. Keaveney, C. S. Adams, and I. G. Hughes,
\href{https://doi.org/10.1103/PhysRevLett.118.253601}{Phys. Rev. Lett. \textbf{118}, 253601 (2017)}.

\bibitem{Science362.446(2018)}
Fabian Ripka, Harald K\"ubler, Robert L\"ow, and Tilman Pfau,
\href{https://doi.org/10.1126/science.aau1949}{Science \textbf{362}, 446-449 (2018)}.

\bibitem{PRA101.023828(2020)}
Tiago J. Arruda, Romain Bachelard, John Weiner, Sebastian Slama, and Philippe W. Courteille,
\href{https://doi.org/10.1103/PhysRevA.101.023828}{Phys. Rev. A \textbf{101}, 023828 (2020)}.

\bibitem{PRL124.063603(2020)}
Sebastian Wolf, Stefan Richter, Joachim von Zanthier, and Ferdinand Schmidt-Kaler,
\href{https://doi.org/10.1103/PhysRevLett.124.063603}{Phys. Rev. Lett. \textbf{124}, 063603 (2020)}.

\bibitem{PRA100.033833(2019)}
P. Weiss, A. Cipris, M. O. Ara\'ujo, R. Kaiser, and W. Guerin,
\href{https://doi.org/10.1103/PhysRevA.100.033833}{Phys. Rev. A \textbf{100}, 033833 (2019)}.

\bibitem{PRL122.183203(2019)}
N. Cherroret, M. Hemmerling, V. Nador, J. T. M. Walraven, and R. Kaiser,
\href{https://doi.org/10.1103/PhysRevLett.122.183203}{Phys. Rev. Lett. \textbf{122}, 183203 (2019)}.

\bibitem{PhysRev93.99(1954)}
R. H.~Dicke, 
\href{https://doi.org/10.1103/PhysRev.93.99}{Phys. Rev. \textbf{93}, 99 (1954)}.

\bibitem{PRA15.1613(1977)}
G. S. Agarwal, A. C. Brown, L. M. Narducci, and G. Vetri,
\href{https://doi.org/10.1103/PhysRevA.15.1613}{Phys. Rev. A \textbf{15}, 1613 (1977)}.

\bibitem{PRA64.063801(2001)}
C. Skornia, J. von Zanthier, G. S. Agarwal, E. Werner, and H. Walther, 
\href{https://doi.org/10.1103/PhysRevA.64.063801}{Phys. Rev. A \textbf{64}, 063801 (2001)}.

\bibitem{PRA19.1132(1979)} 
Helen S. Freedhoff,
\href{https://doi.org/10.1103/PhysRevA.19.1132}{Phys. Rev. A \textbf{19}, 1132 (1979)}.

\bibitem{OptActa29.265(1982)}
Thomas Richter,
\href{https://doi.org/10.1080/713820845}{Opt. Acta, \textbf{29}:3, 265-273 (1982)}.

\bibitem{PRA25.1528(1982)}
R. D. Griffin and S. M. Harris,
\href{https://doi.org/10.1103/PhysRevA.25.1528}{Phys. Rev. A \textbf{25}, 1528 (1982)}.

\bibitem{PRA29.2004(1984)}
Z. Ficek, R. Tana\'s, and S. Kielich,
\href{https://doi.org/10.1103/PhysRevA.29.2004}{Phys. Rev. A \textbf{29}, 2004 (1984)}.

\bibitem{PRA52.636(1995)}
T. G. Rudolph, Z. Ficek, and B. J. Dalton,
\href{https://doi.org/10.1103/PhysRevA.52.636}{Phys. Rev. A \textbf{52}, 636 (1995)}.

\bibitem{WillThesis}
W. J. Hamlyn, {\it A new platform for atom-light interactions on the nano-scale}. Doctoral thesis, Durham University (2020)
\href{http://etheses.dur.ac.uk/13565}{http://etheses.dur.ac.uk/13565}

\bibitem{OptLet40.4289(2015)}
Daniel J. Whiting, Erwan Bimbard, James Keaveney, Mark A. Zentile, Charles S. Adams, and Ifan G. Hughes, 
\href{https://www.osapublishing.org/ol/abstract.cfm?uri=ol-40-18-4289}{Optics Letters \textbf{40}, pp. 4289-4292 (2015).}

\bibitem{PhysRep372.369(2002)}
Z. Ficek and R. Tana\'s,
\href{https://doi.org/10.1016/S0370-1573(02)00368-X}{Phys. Rep. \textbf{372}, 369-443 (2002)}.


\bibitem{BookCarmichael}
Howard J. Carmichael, Statistical Methods in Quantum Optics 1 (Springer, Berlin, Heidelberg, 2002).

\bibitem{JPhysB50.014004(2017)}
Ryan Jones, Reece Saint and Beatriz Olmos,
\href{https://doi.org/10.1088/1361-6455/50/1/014004}{J. Phys. B: At. Mol. Opt. Phys. \textbf{50}, 014004 (2017).}

\bibitem{JPhysB16.2677(1983)}
G. Nienhuis, 
\href{https://doi.org/10.1088/0022-3700/16/15/011}{J. Phys. B.: At. Mol. Phys. \textbf{16}, 2677 (1983).}

\bibitem{JPhysB20.4915(1987)}
J. D. Cresser,
\href{https://doi.org/10.1088/0022-3700/20/18/027}{ J. Phys. B.: At. Mol. Phys. \textbf{20}, 4915 (1987).}

\bibitem{JKeaveneyThesis}
J. Keaveney, Collective Atom Light Interactions in Dense Atomic Vapours (Springer, Berlin, Heidelberg, 2014).

\bibitem{RubidiumData}
D. A. Steck, 
\href{ http://steck.us/alkalidata}{Rubidium 85 D Line Data (2009)}.

\bibitem{RevModPhys15.1(1943)}
S. Chandrasekhar,
\href{https://doi.org/10.1103/RevModPhys.15.1}{Rev. Mod. Phys. \textbf{15}, 1, (1943)}.

\bibitem{PRL112.113603(2014)}
Juha Javanainen, Janne Ruostekoski, Yi Li, and Sung-Mi Yoo,
\href{https://link.aps.org/doi/10.1103/PhysRevLett.112.113603}{Phys. Rev. Lett. \textbf{112}, 113603 (2014).}

\bibitem{PRA96.033835(2017)}
Juha Javanainen, Janne Ruostekoski, Yi Li, and Sung-Mi Yoo,
\href{https://link.aps.org/doi/10.1103/PhysRevA.96.033835}{Phys. Rev. A \textbf{96}, 033835 (2017).}

\bibitem{PRA58.4133(1998)}
Almut Beige and Gerhard C. Hegerfeldt,
\href{https://doi.org/10.1103/PhysRevA.58.4133}{Phys. Rev. A \textbf{58}, 4133 (1998)}.

\bibitem{PRL125.073601(2020)}
A. Cidrim, T. S. do Espirito Santo, J. Schachenmayer, R. Kaiser, and R. Bachelard,
\href{https://link.aps.org/doi/10.1103/PhysRevLett.125.073601}{Phys. Rev. Lett. \textbf{125}, 073601 (2020)}

\bibitem{Tom}
T. F. Cutler, W. J. Hamlyn, J. Renger, K. A. Whittaker, D.  Pizzey, I. G. Hughes, V. Sandoghdar, and C. S. Adams, 
\href{https://link.aps.org/doi/10.1103/PhysRevApplied.14.034054}{Phys. Rev. Applied \textbf{14}, 034054 (2020)}.

\bibitem{KWhittaker2015}
K.  A. Whittaker, J. Keaveney, I. G. Hughes, A. Sargysyan, D. Sarkisyan, B. Gmeiner, V. Sandoghdar, and C. S. Adams, 
\href{http://dx.doi.org/10.1088/1742-6596/635/12/122006}{J. Phys. Conf. Ser. {\bf 635}, 122006 (2015).}

\bibitem{KWhittakerThesis}
K. Whittaker, {\it Construction and characterisation of ultra-thin alkali-metal vapour cells}. Doctoral thesis, Durham University (2017)
\href{http://etheses.dur.ac.uk/12112/}{http://etheses.dur.ac.uk/12112/}

\bibitem{dataDOI}
{\it Data is available through Durham University data management}
\href{http://doi.org/10.15128/r2cz30ps686}{http://doi.org/10.15128/r2cz30ps686}

\bibitem{OstermannThesis}
L. Ostermann, {\it Collective radiation of coupled atomic dipoles and the precise measurement of time}. Doctoral thesis, University of Innsbruck (2016)
\href{https://diglib.uibk.ac.at/ulbtirolhs/content/titleinfo/1371461}{https://diglib.uibk.ac.at/ulbtirolhs/content/titleinfo/1371461}


\end{thebibliography}
\end{document}